\documentclass [12pt,preprint]{aastex}
\usepackage{epsfig}
\begin{document}
\voffset-.5cm
\newcommand{\gsim}{\hbox{\rlap{$^>$}$_\sim$}}
\newcommand{\lsim}{\hbox{\rlap{$^<$}$_\sim$}}

\title{`Jet breaks' and `missing breaks' in the X-Ray afterglow\\ 
                  of Gamma Ray Bursts }

\author{Shlomo Dado\altaffilmark{1}, Arnon Dar\altaffilmark{2}
and A. De  R\'ujula\altaffilmark{3}}

\altaffiltext{1}{dado@phep3.technion.ac.il\\ 
Physics Department and Space Research Institute, 
Technion, Haifa 32000, Israel}
\altaffiltext{2}{arnon@physics.technion.ac.il; arnon.dar@cern.ch \\ 
Physics Department and Space Research Institute,
Technion, Haifa 32000, Israel\\
Theory Unit, CERN, 1211 Geneva 23, Switzerland }
\altaffiltext{3}{alvaro.derujula@cern.ch\\
 Theory Unit, CERN,1211 Geneva 23, Switzerland and\\
Physics Department, Boston University, USA}

\begin{abstract}

The X-ray afterglows (AGs) of Gamma-Ray Bursts (GRBs) and X-Ray Flashes
(XRFs) have, after the fast decline phase of their prompt emission,
a temporal behaviour varying between two extremes. A large fraction of these AGs
has a `canonical' light curve which, after an initial shallow-decay
`plateau' phase,  `breaks smoothly' into a fast power-law decline.
Very energetic GRBs, contrariwise, appear not to have a
`break', their AG declines like a power law from the start of the observations.
Breaks and `missing breaks' are  intimately related to the
geometry and deceleration of the jets responsible for GRBs.
In the frame of the `cannonball' (CB) model of GRBs and XRFs,
we analyze the cited extreme behaviours (canonical and pure power law)
and  intermediate cases spanning the observed range of
X-ray AG shapes. We show that the entire panoply of X-ray light-curve
shapes --measured with Swift and other satellites-- are as anticipated
in the CB model. We test the expected correlations
between the AG's shape  and the peak- and isotropic energies
of the prompt radiation, strengthening a  simple conclusion
of the  analysis of AG shapes: in energetic GRBs the break is
not truly `missing', it is hidden under the tail of the prompt emission,
or it occurs too early to be recorded.
We also verify that the spectral index of the unabsorbed AGs
and the temporal index of their late power-law decline
differ by half a unit, as predicted.

\end{abstract}

\section{Introduction and r\'esum\'e}

The isotropic distribution of gamma ray bursts (GRBs) in the sky and their 
number distribution as function of intensity, measured with the 
BATSE instrument aboard the Compton Gamma Ray Observatory, 
provided the first observational evidence that gamma ray 
bursts (GRBs) originate at large cosmological distances (Meegan et al.~1992).  
Moreover, the rapid variation of their light curves (Bahat et al.~1992) 
indicated that their huge energy is emitted from a very small volume. 
In the original fireball (FB) model of GRBs
(e.g. Paczynski~1986;  Goodman~1986; Rees \& 
M\'esz\'aros~1992) the emission was spherically symmetric. 
The implied isotropic energy release of GRBs in $\gamma$-rays often
exceeded $M_\odot\, c^2$, creating an `energy crisis'. Indeed, such a
mighty, abrupt, compact, and $\gamma$-ray-efficient source was 
unforeseen.

A simple solution to this puzzle was suggested by Shaviv and Dar 
(1995): the $\gamma$-ray emission is narrowly 
collimated by the relativistic motion of their {\it jetted} source, which is seen
when it points closely enough to the observer. In this view, GRBs are not 
produced by fireballs, but by inverse Compton scattering of light by highly 
relativistic jets of ordinary matter, ejected in violent stellar processes 
such as supernova explosions, mergers of neutron stars 
(Paczynski~1986; Goodman, Dar \& Nussinov~1987; Dar et al.~1992), or the
direct collapse of massive stars to black holes without a supernova 
(Woosley 1993). 

The sky localization of GRBs by BeppoSAX (Costa et al.~1997) led to the
discovery (Groot et al.~1997; van Paradijs et al.~1997) of their optical
afterglows (AGs) and their host galaxies (Sahu et al.~1997)  which were
used to extract their cosmological redshifts (Metzger et al.~1997). The
AGs seemed to follow an achromatic power-law decline, as expected from a
highly relativistic expanding fireball that drives a blast wave into the
circumburst environment (e.g., M\'esz\'aros \& Rees~1997). 
This prediction of the spherical
fireball model (see e.g., Piran~1999) being independent of the assumption 
of spherical symmetry,
it was also argued that the AGs, like the GRBs themselves, are produced by
narrowly collimated jets (Dar~1997, 1998).

The concept of jets was incorporated into the FB model by
the substitution of its spherical shells by conical sections thereof, the
mechanism for the $\gamma$-ray emission still being synchrotron radiation
from shock-accelerated $e^+\,e^-$ pairs in a baryon-poor material (see,
e.g.~Piran~1999,~2000; M\'esz\'aros~2002 and references therein), in
spite of the difficulties that such a radiation mechanism encounters
(Ghisellini et al.~2000).

An elegant and simple way to distinguish between a conical jet and a spherical fireball 
was suggested by Rhoads (1997):  the AG of a decelerating conical jet will show
an achromatic steepening --a {\it jet break}-- in its power-law decline when
the relativistic beaming angle of its radiation becomes larger than the
opening angle of the jet. Soon afterwards, better sampled data on
the optical afterglow of GRBs showed the existence of what appeared to be
such achromatic jet breaks (Harrison et al.~1999; Stanek et al.~1999), 
and the spherical FB model was
modified into a {\it collimated fireball} model (e.g.~Piran, Sari
 \& Halpern~1999). In this model GRB pulses are produced by synchrotron
radiation from the collision between conical sections of shells. The
collision of the ensemble of shells with the interstellar matter (ISM)
generates the AG by synchrotron radiation from the forward shock
propagating in the ISM, and/or from the backwards shock within the merged
shells.  Rhoads (1999) and Sari, Piran and Halpern (1999) derived a
relation between the opening angle of the conical jet and the time of the
jet break. This relation has been applied extensively to the pre-Swift
data to infer the opening angle of the conical jet and to determine the
`true' energy of GRBs, posited to be an approximate standard candle (Frail
et al.~2001).
  
Since the launch of Swift the above generally-accepted `standard' paradigm 
has been challenged, due to the absence of breaks in the AGs of many GRBs 
(Panaitescu et al.~2006;  Burrows and Racusin 2006), to the chromatic 
behaviour of the AG of other GRBs having the alleged `jet break' (Stanek 
et al.~1999, Harrison et al.~1999), and to the failure of the Frail 
relation (Frail et al.~2001) in many Swift GRBs (Kocevski \& Butler~2007). 
 In the fireball model, the jet breaks need not be sharp; they are 
often parametrized with a varying smoothness (Stanek et al.~1999).
Allowing for such breaks, Covino et al. (2006) could not identify a 
Swift GRB with a fully achromatic break. Liang et al.~(2007) have 
extended this study, and analyzed the Swift X-ray data for the 
179 GRBs detected between January 2005 and January 2007 
and the optical AGs of 57 pre- and post-Swift GRBs. They found 
that not a single burst satisfies all the criteria of a jet break.  
This brings us fully into the question of the nature and properties of the 
jets responsible for GRBs and their AGs and, more specifically in this 
paper, to the understanding of `breaks' and `missing breaks'.

An alternative to the fireball scenario is offered by the `cannonball' 
(CB) model of GRBs [Dar \& De R\'ujula~2000a, 2004, hereafter DD2000a, 
2004; Dado, Dar \& De R\'ujula (hereafter DDD)~2002; for a recent review 
see De R\'ujula~2007]. In this model {\it long-duration} GRBs and their 
AGs are produced by bipolar jets of CBs (Shaviv \& Dar~1995; Dar \& 
Plaga~1999), ejected in~{\it ordinary core-collapse} supernova (SN) 
explosions as matter is accreted onto the newly-formed compact object (De 
R\'ujula 1987). The `cannon-balls' are made of {\it ordinary-matter 
plasma}. The $\gamma$-rays of a single pulse of a GRB are produced as a CB 
coasts through the SN {\it glory} --the initial SN light, scattered away 
from the radial direction by the `wind': the ejecta puffed by the 
progenitor star in a succession of pre-SN flares. The electrons enclosed 
in the CB raise the glory's photons to GRB energies by {\it inverse 
Compton scattering (ICS)}. As a CB coasts through the glory, the 
distribution of the glory's light becomes increasingly radial and its 
density decreases rapidly. Consequently, the energy of the up-scattered 
photons is continuously shifted to lower energies and their number 
decreases swiftly, resulting in a fast softening and decline of the prompt 
emission (DD2004, 2007a,b). In the CB model, the AG of a GRB is due to 
{\it synchrotron radiation (SR)} from swept-in ISM electrons spiraling in 
the CB's enclosed turbulent magnetic field, generated by the intercepted 
ISM nuclei and electrons (DDD2002). At X-ray energies, the SR afterglow 
begins to dominate the ICS prompt emission only during the fast-decline 
phase of the latter (DDD2006).

In the CB model, the {\it beau r\^ole} in the understanding of GRBs 
is played by the Doppler factor, $\delta(t)$, relating times,
energies and fluxes in a CB's rest system to those in the observer's system. 
Its  form
in terms of the observer's angle $\theta$ (relative to the CB's direction
of motion) and the time-dependent Lorentz factor, $\gamma(t)$, of a CB, is:
\begin{equation}
\delta(t)={1\over \gamma(t)\,
[1-\beta(t)\, \cos\theta]}\approx {2\, \gamma(t)\over
1+[\gamma(t)\, \theta]^2}\; ,
\label{delta}
\end{equation}
where the approximation is excellent for $\gamma\!\gg\! 1$ and 
$\theta\!\ll\! 1$.
The decrease of $\gamma(t)$ with time, as a CB encounters the particles
of the ISM, is calculable on grounds of energy-momentum conservation
(DDD2002, Dar \& De R\'ujula 2006, thereafter DD2006). The 
energy-integrated energy flux 
of the AG of a GRB,
is $\propto\!\delta^3$. Let $\gamma_0\!\equiv\!\gamma(0)$.
Consider a CB that is observed almost on axis,
so that $\theta\,\gamma_0\!<\! 1$: the observer is
{\it ab initio} within the opening cone of the relativistically beamed radiation.
As $\gamma(t)$ decreases, $\delta(t)$ monotonically decreases and so does
the observed AG. Consider the same CB, viewed by an observer at a
much larger angle, so that $\theta\,\gamma_0$ is `a few'.
 As $\gamma(t)$ decreases, $\delta(t)$ in Eq.~(\ref{delta}) {\it increases},
 reflecting the fact that the characteristic opening angle
 of the radiation, $1/\gamma(t)$,
 is reaching the observer's direction. Past the point $\gamma\,\theta\!\sim\!1$,
the decrease of $\delta(t)$ is monotonic,
as in the first case we considered. The AG radiation parallels again the behaviour
of $\delta(t)$. For observers of the same GRB from different angles, as
 $\theta$ increases at fixed $\gamma(t)$, the AG's flux decreases.
All these simple facts, supported by the corresponding explicit derivations,
are reflected in Fig.~\ref{f1}a, which we have copied from DDD2002, as it
foretells the progressive variety of AG shapes to be studied here. 

There is more to Fig.~\ref{f1}a than what we said. The
Lorentz factor $\gamma(t)$ of a CB only begins to change significantly, in a
calculable manner, when the
increase in its mass --induced by the energy influx of the swept-in ISM particles--
becomes comparable to the CB's initial mass. This happens, as we shall review,
at a time $t_b\!\propto\![1+2\,\theta^2\,\gamma_0^2]/\gamma_0^3$.
 At fixed $\gamma(t)$, as reflected in 
Fig.~\ref{f1}a, a larger $\theta$ entails a larger $t_b$.
This achromatic `deceleration bend' at $t\!=\!t_b$, we believe,
was often interpreted in FB models as a putative 
jet break.

Naturally, the values of $\gamma_0$ and $\delta_0$ of a given CB also affect 
the properties of its prompt ICS-dominated radiation (we are presenting this
introductory discussion as if there was a single CB generating the prompt
and AG radiations, a simplification to be undone when needed). 
In the CB model the ICS-dictated $(\theta,\,\gamma_0)$
dependences of a CB's isotropic energy, peak energy
and peak luminosity are $E_{\rm iso}\!\propto\!\delta_0^3$,
$E_p\!\propto\!\gamma_0\,\delta_0$ and $L_p\!\propto\!\delta_0^4$ (DD2000b).
The conditions for these quantities to be relatively large (a relatively
small $\theta$ or a large $\gamma_0$) are the ones leading to a luminous AG
with a small $t_b$. The basis for one of these expected correlations, studied before 
in detail in DDD2007c, is illustrated in Fig.~\ref{f1}b.

If the deceleration bend at time $t_b$ takes place {\it after} the fast-decline phase
of the prompt emission, it is observable, and the unabsorbed X-ray light
curve is canonical (DDD2002).
In these cases, there is a `break'.
If $t_b$ takes place earlier, it is hidden under the prompt
emission, and only the tail of the canonical behaviour, namely the `late'
power-law decline of the unabsorbed synchrotron afterglow, is observable.
In these cases, the break is missing. The transition from long-plateau,
clearly `broken' AGs, to power-law like `unbroken' AGs should be
anticorrelated with the trend from under-`energetic' to over-energetic GRBs.

In the CB model the late-time spectral energy density $F_\nu$ of 
the X-ray and optical AG tends to
a time and energy-dependence 
$\propto\!t^{-(p+1)/2}\,\nu^{-p/2}$, with $p$ the spectral index of the
electrons accelerated within a CB and cooled by the emission of the
very SR seen as the AG.
A  prediction that we have not emphasized before is
that the temporal power decline should be, GRB by GRB, half a unit steeper
than the spectral decline. 

In DDD2006, 2007a we have demonstrated that the most common light curves 
of the X-ray AG of GRBs are well described by the CB model. We have also 
explained there the various origins of the chromatic behaviours of AGs.  
In DDD2007b we have focused on the fast decline phase of the prompt 
emission and we have demonstrated that the rapid spectral evolution 
observed during this phase is also as expected in the CB model. In 
DDD2007c we have shown, for large ensembles of GRBs, how the observed 
correlations between $E_{\rm iso}$, $E_p$ [Dar \& De R\'ujula 2000b 
(DD2000b), Amati et al.~2002], $L_p$, and other prompt observables (pulse 
rise-time, lag-time and variability) follow mainly from the same simple 
geometrical considerations --that we have reviewed above-- on the 
case-by-case variability of the Doppler factor. In the CB model, XRFs are 
simply GRBs seen at relatively large $\theta$ (DDD2004a), even the 
particularly interesting XRF 060218 is in no way exceptional (DDD2007a).

In this paper we
focus on the shape of the light curves of the X-ray afterglow of GRBs,
with and without breaks, measured with the X-ray telescope (XRT) aboard
Swift. We show that the shapes of the X-ray light curves of GRBs and XRFs
predicted in Fig.~\ref{f1}a, and the
correlation between  $t_b$ and $E_{\rm iso}$ illustrated Fig.~\ref{f1}b
(and the consequent apparent presence or absence of breaks in
the AG) agree with the CB-model's expectations. We also analize the
$(t_b,\,E_p)$ correlation on the same light. Finally, we investigate
the relation between the temporal power-law index of the post-break
decline and the photon spectral index, reaching satisfactory results.
To do all this, we investigate 16 GRBs chosen to reflect the
full span of the question of the presence or absence of breaks.
The selected GRBs range from the faintest known GRB (980425, of
supernova-association fame), which also has the most pronounced plateau
and the latest break time, to the brightest Swift GRB (061007), with the 
most
luminous and longest-observed unbroken power-law X-ray AG.

\section{The afterglow of a decelerating CB}

In the CB model, the mechanism for the emission of the prompt radiation
of GRBs and XRFs is inverse Compton scattering. 
The temporal and spectral properties of the prompt phase,
including its fast decline, are summarized in a `master formula' 
(DD2004) that we have already contrasted with Swift data (DDD2006, 2007a,b,c,d). We 
shall not  repeat it here ---as our emphasis in the current study is 
on breaks in the X-ray light curves of GRB afterglows--- though we shall 
use it to describe the fading of the prompt emission until the take-over by 
the synchrotron-AG emission, and the occasional late X-ray
flares. Neither do we discuss here the 
optical AGs (DDD2007a). 
The extinction in the optical- and, more so, in the radio- domain
(within the CBs, in the circumburst environment, in the ISM of the 
host galaxy and ours, and in the intergalactic medium) are 
 difficult to model as reliably as the X-ray extinction.
 We shall see once again that the X-ray light 
curves (corrected for extinction) carry clear and direct information on the radiation 
mechanisms that dominate the prompt emission and the AG phase (ICS and SR,
respectively, in the CB model).

During the initial phase of $\gamma$-ray emission in a GRB,
the Lorentz factor $\gamma$ of a CB stays put at its initial value
$\gamma_0\!=\!{\cal{O}}(10^3)$, for the deceleration induced by
the interactions with the ISM has not yet had a significant effect.
The Doppler factor by which the light emitted by a CB is boosted 
in energy is given by Eq.~(\ref{delta}).
Since the emitted light is forward-collimated into a cone of
characteristic opening angle $1/\gamma$,
the boosted energetic radiation is easiest to detect for
$\theta\!=\!{\cal{O}}(1/\gamma_0)$. Thus, typically, 
$\delta_0\!=\!{\cal{O}}(10^3)$.

As a CB ploughs through the ISM, fully ionized by the
preceding $\gamma$ radiation, it gathers and scatters 
the
ISM ions, mainly protons. These encounters are `collisionless'
since, at about the time it becomes transparent to radiation, a CB also
becomes `transparent' to hadronic interactions. As a consequence
of momentum conservation, the scattered and 
re-emitted protons inevitably exert an inwards `pressure' on the CB.
We have assumed that the main effect of this pressure
is to slow the CB's expansion, posited to be relativistic at the emission time.
In the approximation of isotropic re-emission in the CB's rest
frame and a constant ISM density $n$, one then finds that, typically
within minutes of observer's time $t$, a CB reaches a roughly
`coasting' radius, $R\!=\!{\cal{O}}(10^{14}\,\rm cm)$, which increases
slowly until the CB finally stops and blows up  (DD2006).
Up to the end of the coasting phase,
and in a constant density ISM, $\gamma(t)$
obeys (DDD2002):
\begin{eqnarray}
&&({\gamma_0/ \gamma})^{3+\kappa}+
(3-\kappa)\,\theta^2\,\gamma_0^2\,(\gamma_0/\gamma)^{1+\kappa}
= 1+(3-\kappa)\,\theta^2\,\gamma_0^2+t/t_0\,,
\nonumber\\
&&t_0 = {(1+z)\, N_{_{\rm B}}\over
(6+2\kappa)\,c\, n\,\pi\, R^2\,\gamma_0^3}\,,
\label{deceleration}
\end{eqnarray}
where $\kappa\!=\!1$ if the ISM particles re-emitted fast by the CB are a
small fraction of the flux of the intercepted ones. In the opposite limit,
$\kappa\!=\!0$. In the CB model of cosmic rays (DD2006),
the observed spectrum strongly favours $\kappa\!=\!1$, used here in our
fits. We have also concluded  from previous analysis of Swift
X-ray data that $\kappa\!\approx\!1$ is the right 
choice.

As indicated by first-principle calculations of the relativistic merger of two
plasmas (Frederiksen et al.~2004), the ISM ions continuously impinging on 
a CB generate 
within it turbulent magnetic fields, which we assume to be
in approximate energy equipartition with the energy of the
intercepted ISM, $B\approx \sqrt{2\, \pi\, n\, m_p\, c^2}\, \gamma$.
In this field, the intercepted
electrons emit synchrotron radiation. The SR, isotropic in the CB's
rest frame, has a characteristic frequency, $\nu_b(t)$,
the typical frequency radiated by the
electrons that enter a CB at time $t$ with a relative Lorentz
factor $\gamma(t)$. In the observer's frame:
\begin{equation}
\nu_b(t)\simeq  {\nu_0 \over 1+z}\,
{[\gamma(t)]^3\, \delta(t)\over 10^{12}}\,
\left[{n\over 10^{-1}\;\rm cm^3}\right]^{1/2}
{\rm Hz}.
\label{nub}
\end{equation}
where $\nu_0\!\sim\! 8.5\times 10^{16}\, \rm Hz \simeq 354\, eV$. 
The spectral energy density of the SR
from a single CB at a luminosity distance $D_L$  is given by (DDD2003a):
\begin{equation}
F_\nu \simeq {\eta\,  \pi\, R^2\,n\, m_e\, c^3\,
\gamma(t)^2\, \delta(t)^4\, A(\nu,t)\,
\over 4\,\pi\, D_L^2\,\nu_b(t)}\;{p-2\over p-1}\;
\left[{\nu\over\nu_b(t)}\right]^{-1/2}\,
\left[1 + {\nu\over\nu_b(t)}\right]^{-(p-1)/2}\,,
\label{Fnu}
\end{equation}
where $p\sim 2.2$ is the typical spectral 
index\footnote{The normalization in Eq.~(\ref{nub}) is only correct for $p\!>\!2$,
for otherwise the norm diverges. The cutoffs for the $\nu$ distribution are
time-dependent,
dictated by the acceleration and SR times of electrons  and their `Larmor'
limit. The discussion of these processes being complex (DDD2003a, DD2006),
we shall satisfy ourselves here with the statement that for
$p\!\leq \!2$ the AG's normalization is not predicted.}
of the Fermi accelerated
electrons, $\eta\!\approx\!1$ is the fraction of the impinging ISM 
electron
energy that is synchrotron re-radiated by the CB, and $A(\nu, t)$ is
the  attenuation of photons of observed frequency $\nu$ along the
line of sight through the CB, the host galaxy (HG), the intergalactic medium        
(IGM)
and the Milky Way (MW):
\begin{equation}
A(\nu, t) = {\rm 
exp[-\,\tau_\nu(CB)-\tau_\nu(HG)-\tau_\nu(IGM)-\tau_\nu(MW)].}
\label{attenuation}
\end{equation}
The opacity
$\tau_\nu\rm (CB)$ at very early times, during the fast-expansion phase of the CB, may
 strongly depend  on time and frequency. The opacity of the circumburst medium
 [$\tau_\nu\rm (HG)$ at early times] is affected by the GRB and could also be $t$-
and $\nu$-dependent.  The opacities $\tau_\nu\rm (HG)$
and $\tau_\nu\rm (IGM)$ should be 
functions of $t$ and $\nu$, for the line of sight to the CBs varies
 during the AG observations, due to the hyperluminal motion
of CBs. These facts, the different $(t,\,\nu)$ dependences
of the ICS and SR emissions, and the dependence of the
synchrotron AG on $\nu_b(t)$, are responsible for the complex observed
chromatic behaviour of the AGs. To a fair approximation, though, the
deceleration bend, if occurring late enough, is achromatic from X-ray
energies to the optical domain (DDD2002) but not as far as radio (DDD2003a).

The Swift X-ray bands are above the characteristic frequency $\nu_b$ in Eq.~(\ref{nub})
at all times.
It then follows from Eq.~(\ref{Fnu}) that
 the {\it unabsorbed} X-ray spectral energy density has the form:
\begin{equation}
F_\nu \propto R^2\, n^{(p+2)/4}\,
\gamma^{(3p-2)/2}\, \delta^{(p+6)/2}\,  \nu^{-p/2}=
R^2\, n^{\Gamma/2}\,
\gamma^{3\,\Gamma-4}\, \delta^{\Gamma+2}\, \nu^{-\Gamma+1}\, ,
\label{Fnux}
\end{equation}
where we have used the customary notation $dN_{\gamma}/dE\!\approx\!E^{-\Gamma}$.
 
\section{Breaks, missing breaks, and the AG's asymptotic power decline}

The functions $\delta(t)/\delta_0$ and $\gamma(t)/\gamma_0$
of Eqs.~(\ref{delta},\ref{deceleration}) evolve slowly, up
until a time:
\begin{eqnarray}
\!\!\!\!\!\!\!\!&&t_b=[1+2\, \theta^2\,\gamma_0^2]\;t_0\nonumber\\
\!\!\!\!\!\!\!\!&&\approx
(130\,{\rm s})\, [1+2\,\gamma_0^2\, \theta^2]\,(1+z)
\left[{\gamma_0\over 10^3}\right]^{-3}\,
\left[{n\over 10^{-1}\, {\rm cm}^{-3}}\right]^{-1}
\left[{R\over 10^{14}\,{\rm cm}}\right]^{-2}
\left[{N_{_{\rm B}}\over 10^{50}}\right] \!,
\label{tbreaks}
\end{eqnarray}
where we scaled the result to typical CB-model values of $R$ and 
a CB's baryon number, $N_{_{\rm B}}$. The
combination of the parameters $n$, $R$ and $N_{_{\rm B}}$ appearing in
Eq.~(\ref{tbreaks}) is best constrained by the excellent
X-ray observations discussed here.
Our previous results on optical and radio AGs (for fixed $R$ and $N_{_{\rm B}}$)
favoured 10 times a smaller $n$ at the much larger sampled times, 
not an inconsistency, since a CB  travels for $\sim\!\gamma\,\delta$ light-days
in one day of GRB data. We have chosen
to normalize $n$ as in Eq.~(\ref{tbreaks}), rather than to reproduce long discussions
on the distributions of CB-model parameters (e.g.~De R\'ujula~2007).

The quantity $t_b$ in Eq.~(\ref{tbreaks}) characterizes the {\it deceleration bend-time}
of the CB model; Eq.~(\ref{deceleration}) for $\gamma(t,t_0,\theta,\gamma_0)$
describes the gradual character of this `break'.
At later times  Eq.~(\ref{deceleration})
implies that $\gamma\to\gamma_0\,(t/t_0)^{-1/4}$, and Eq.~(\ref{delta})
that $\delta\to 2\,\gamma$. Thus, at $t\! \gg \!t_b$,
Eq.~(\ref{Fnux}) yields:  
\begin{eqnarray}
F_\nu(t)&\propto& t^{-1/2-p/2}\,\nu^{-p/2}= t^{-\Gamma+1/2}\, \nu^{-\Gamma+1},
\nonumber\\
p&=&2\,(\Gamma -1)
\label{Asymptotic}
\end{eqnarray}
with, as announced, a power decay in time half a unit steeper than in
frequency.

\section{The prompt observables} 

In the CB model, the peak energy of GRBs satisfies:
\begin{equation}
 E_p\simeq {1\over 2}\;{\gamma_0\, \delta_0 \over 1+z}\; \epsilon_g\,,  
\label{ep} 
\end{equation} 
where $\epsilon_g\!\sim\! 1$ eV is the typical
energy of the glory's photons, that of the
associated-supernova early light just 
prior to the ejection of  CBs. 
The isotropic  (or spherical equivalent)
energy  of a GRB is  (DD2004): 
\begin{eqnarray} 
E_{\rm iso} &\simeq& 
{\delta_0^3\, L_{_{\rm SN}}\,N_{_{\rm CB}}\,\beta_s\over 6\, c}\,
                      \sqrt{\sigma_{_{\rm T}}\, N_{_{\rm CB}}\over 4\, \pi}\sim
                      (1.2\! \times\! 10^{53}\;{\rm erg})\,V_E,
                     \nonumber\\
                      V_E&\equiv&{\delta_0^3\over 10^9}\,
{L_{_{\rm SN}}\over L_{_{\rm SN}}^{\rm bw}}\,{N_{_{\rm CB}}\over 2}\;
\beta_s\;\sqrt{ N_{_{\rm CB}}\over 10^{50}}\, ,
\label{eiso} 
\end{eqnarray} 
where $L_{_{\rm SN}}$ is the mean supernova early optical luminosity, 
 $N_{_{\rm CB}}$ is the number of CBs in 
the jet, $\beta_s$ is the comoving early
expansion velocity of a CB (in units of $c/\sqrt{3}$), 
and $\sigma_{_{\rm T}}$ is the Thomson cross section. 
For $\langle N_{_{\rm CB}}\rangle\!=\! 6$  (Schaefer~2007), the early 
SN luminosity required to 
produce the mean isotropic energy, $E_{\rm iso}\!\sim\! 4\!\times\! 10^{53}$ 
erg, of ordinary long GRBs, is 
$L_{_{\rm SN}}^{\rm bw}\!\simeq\! 5\!\times\! 10^{42}\, {\rm erg\, 
s^{-1}}$, the estimated early luminosity of SN1998bw. All quantities in
Eq.~(\ref{eiso}) are normalized to their typical CB-model values.
We have normalized to $N_{_{\rm CB}}\!=\! 2$, an adequate mean number
of prominent X-ray pulses in the subset of GRBs analized here.

The results in Eqs.~(\ref{ep},\ref{eiso}) are based on the assumption that
ICS is the mechanism generating the prompt radiation. They depend
on $\gamma_0$ and $\delta_0$, two parameters also appearing in the
description of the SR afterglow. That is why we shall be able to test the
implied correlations, GRB by GRB, between the shape of the AG and the
energetics of the prompt radiation, the very strong dependence of the
$\delta$ on $\theta$ playing once more the major role.

According to Eqs.~(\ref{delta},\ref{ep},\ref{eiso}), CBs with large $\gamma_0$,
and, more so, small $\theta$, produce the largest values of 
$E_p$ and $E_{\rm iso}$: they generate the brightest GRBs. According to
Eq.~(\ref{tbreaks}), such $(\gamma_0,\theta)$ values entail a small
$t_b$, an expectation that our analysis will validate.
In such cases the deceleration bend or `break' of the synchrotron
AG may take place before the beginning of the XRT observations and/or be
hidden under the prompt Compton emission. According to Eq.~(\ref{Fnux}),
these AGs must be very luminous at early times, and according to
Eq.~(\ref{Asymptotic}), they must be well approximated from starters by
the asymptotic power law behaviour given by Eq.~(\ref{Asymptotic}).  Our
analysis will verify all these  predictions.

\section{Comparison with observations}
 
To date, Swift has detected and localized nearly 300 long GRBs, and for 
most of them it followed their X-ray emission until it faded into the 
background.  Incapable of discussing all of them, we analyze the light 
curves of the X-ray afterglow of a set of GRBs with and `without' jet 
breaks, which represent fairly well the entire spectrum of canonical and 
non-canonical X-ray afterglows of GRBs. They include the most extreme 
cases of canonical and non canonical behaviour (GRB 0980425 and GRB 
061126, respectively), the longest-measured canonical and non canonical 
X-ray light curves (GRB 060729 and GRB 061007) and a variety of light 
curves with and without breaks, with and without superimposed X-ray 
flares. Since many CB-model fits to canonical light curves 
of X-ray afterglows  with  `breaks' 
were included in previous publications (DDD2002, 2006, 2007a,b,d) 
we shall discuss in this paper more cases of GRBs with an approximate power-law 
AG than of GRBs with a canonical AG.

We start the fits to the X-ray light curves
during the transition between the rapid decline phase of the 
ICS-dominated prompt emission to the SR-dominated AG
phase. It suffices to include the ICS contribution of the last
prompt-emission pulse (or the last two), because of an exponential 
factor in the pulse shape that suppresses very fast the relative 
contribution of the earlier pulses by the time the data sample
the later ones (DDD2007a,b). 
For the synchrotron contribution, it usually suffices to 
consider a common emission angle $\theta$ and an average initial Lorentz 
factor $\gamma_0$ for the ensemble of CBs. The ISM density along the CBs' 
trajectories is approximated by a constant. We then fit the entire observations 
of the X-ray AG of the selected GRBs by using the `master formula'
[DD2004, Eq.~(11) of DDD2007a],
for the tail of the ICS prompt emission contribution, and Eq.~(\ref{Fnux}) for 
the SR. Many GRBs have late X-ray flares, which we interpret 
as dying pangs of the engine, that is, the emission of CBs in late episodes of accretion into
the recently collapsed central compact object.
These CBs, whose ICS-generated flares can only be seen on the weak background of a
decaying SR `after'-glow (quote-unquote, since the `AG' is observable {\it before} the late `prompt'
flares), are also modeled with the same master formula.

The calculated shape of the energy-integrated X-ray AG, Eq.~(\ref{Fnux}), 
depends only on three fit parameters. Two of them are the product 
$\gamma_0\, \theta$ and the deceleration-bend time, $t_b$, for an on-axis 
observer, as given in the first line of Eq.~(\ref{deceleration}). They 
determine the deceleration-bend time, $t_b$, observed at a viewing angle 
$\theta$, see Eq.~(\ref{tbreaks}). The third fit parameter is the index 
$p$ in the $\gamma$ and $\delta$ time-dependent factors of 
Eq.~(\ref{Fnux}). Unlike in previous analysis, we let $p$ be a free 
parameter, unrelated to the spectral index $\Gamma$, independently 
extracted by the observers from the shape of the X-ray spectrum. This way 
we shall be able to test explicitly the CB-model prediction implied by 
Eq.~(\ref{Fnux}), $p\!=\! 2\,(\Gamma-1)$, or by its more readable 
asymptotic form, Eq.~(\ref{Asymptotic}). In all the cases we study, but 
two (GRBs 071020 and 050416A), a single CB or an `average' CB suffice to 
describe the AG. The occasional need for two CBs in the AG light-curve 
description is not a novelty. The most notable instance is that of GRB 
030329 (DDD2003b).

A comparison between the observed and predicted light curves of the 16
selected GRBs is shown in Figs.~\ref{f2} to \ref{f5}. When well-measured,
the `break' time, $t_b$, is indicated in the figure by an arrow. 
The best fit values, of $p$, $\gamma_0\, \theta$, and  $t_b$ are listed 
in Table \ref{t1}, along with additional observational information on 
these GRBs (redshift,  peak energy, equivalent isotropic energy,  
the start-time of the Swift XRT observations, the  spectral index 
of the unabsorbed AG, and the $\chi^2$ per degree of freedom of the fits). 

Afterglows which exhibit nearly a pure power-law decline, have a $t_b$ 
smaller than $t_s$, the time after trigger when the XRT started its 
observations of the AG, or a $t_b$ smaller than the time when the 
afterglow became brighter than the tail of the prompt emission. Such AGs 
have a nearly power-law shape, $F_\nu\sim t^{-p/2}$. Their fits, however, 
return upper limits for $\gamma_0\,\theta$ and $t_b$, above which the 
shape of the AG deviates from the data. These limiting values are also 
reported in Table \ref{t1}, but the corresponding limit-$t_b$ location 
will not be shown in the figures in the case-by-case analysis, as it 
generally falls off-limits.

In most cases (including many Swift AGs studied in DDD2006, 2007a,b, but 
not shown here), the CB model produces good fits with reduced $\chi^2$ 
values close to unity. Even if the $\chi^2$ figures are good, we generally 
have refrained in the past from reporting them. One reason is that it is 
easy to obtain an excellent $\chi^2$ for a fit that has many data points, 
but misses some that clearly reflect a significant structure (such as a 
supernova, see DDD2002), or is, even within errors, systematically above 
the data in one region and below it in another. For that reason, and the 
occasional local scatter of the data, we consider the eye to be a better 
judge than any statistical measure. We comment on the $\chi^2$
values when they are `bad'.

The values of $\theta\,\gamma_0$, $t_b$ (or $t_0$), and $p$ returned by 
our fits and reported in Table \ref{t1} have formal errors of a few 
percent. The error-correlation matrix has relatively small off-diagonal 
elements. The reduced-$\chi^2$ values are very close to unity, once the 
occasional flares are taken into account, to reveal the presence of a 
smoother SR background. One reason for all this is that $t_0$ sets the 
overall time scale, $\theta\,\gamma_0$ determines the shape of the bend, 
and $p$ is sensitive to the whole SR light curve, playing a major role in 
its power-law tail. This means that when a light curve is well sampled 
(over orders of magnitude in flux and time), the fit is very sensitive to 
its parameters. Naturally, the results depend also on the deceleration 
law, Eq.~(\ref{deceleration}), meant to be an approximation. Therefore the 
extracted parameters have `systematic' errors reflecting the approximate 
nature of Eq.~(\ref{deceleration}). We can argue explicitly why the 
approximation should be better than it looks at first sight, even case by 
case (on average, and independently, it leads to the correct spectrum of 
cosmic rays from non-relativistic energies up to the `knee' at some 
$2\times 10^6$ GeV, DD2006). The continuation of this rather formal 
argument on errors would take us well beyond the scope of this paper.

\subsection*{Case studies}

\begin{figure}[]
\centering
\vspace{-1cm}
\vbox{
\hbox{
\epsfig{file=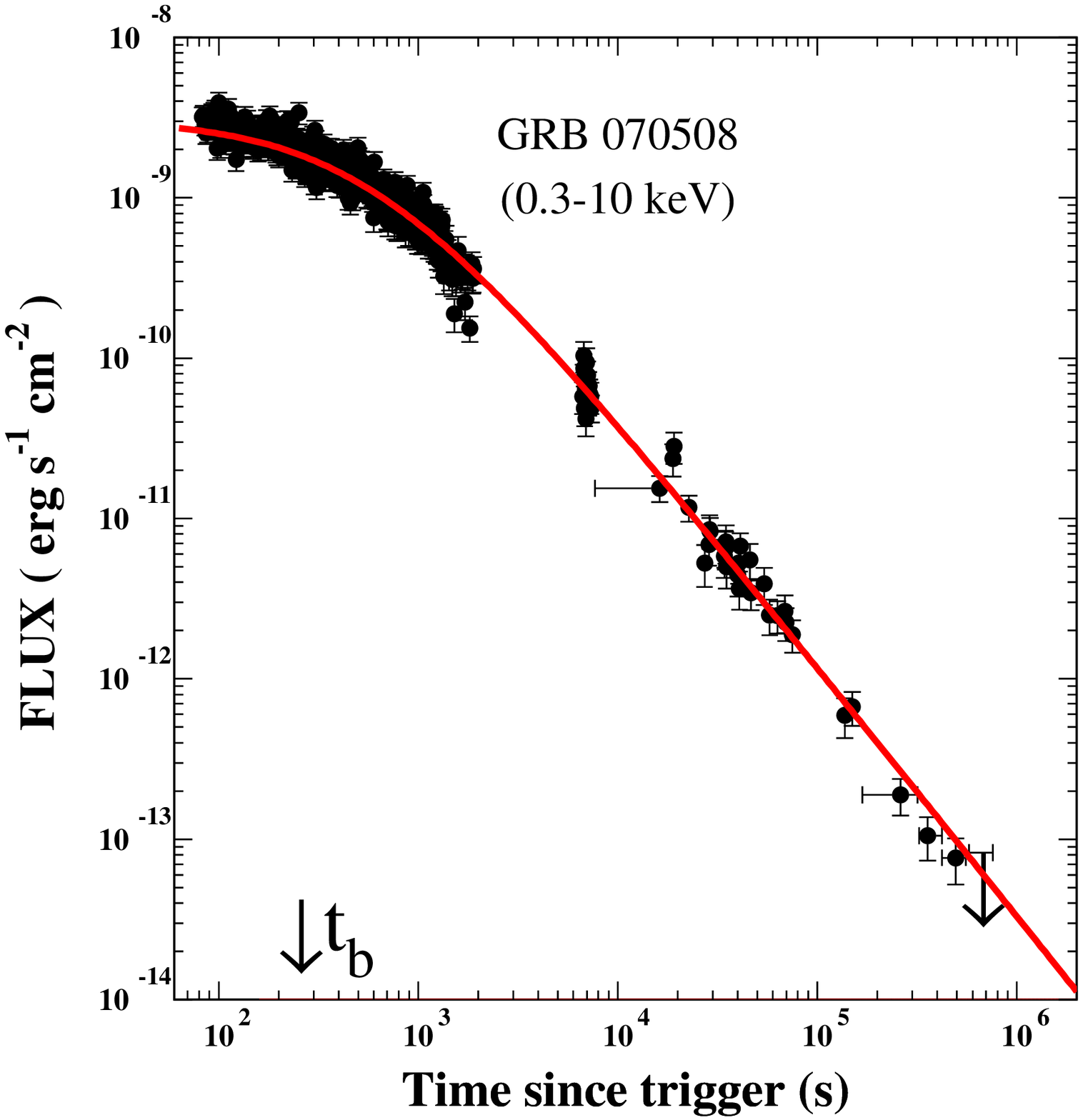,width=8.0cm} 
\epsfig{file=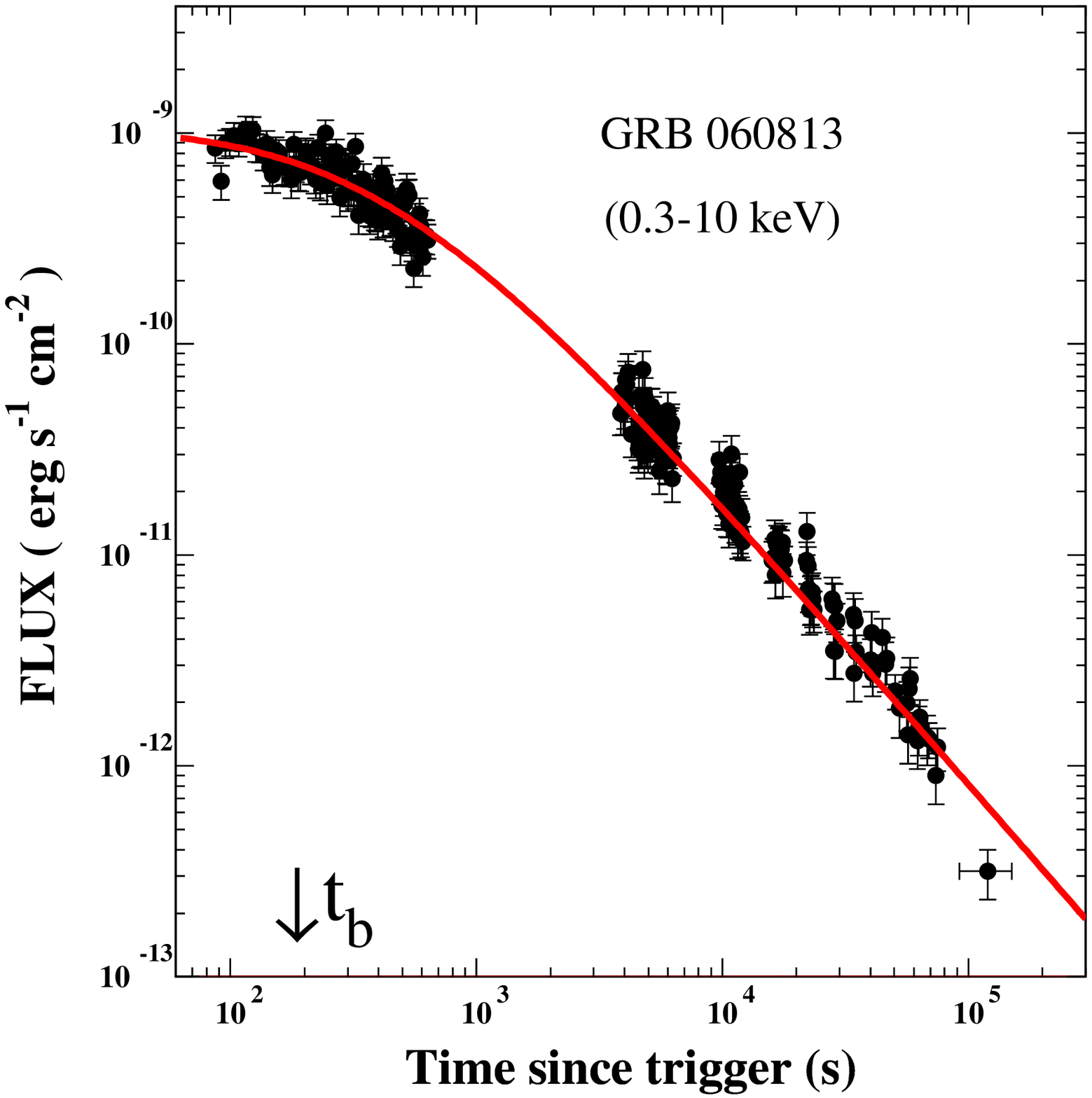,width=8.0cm}
}}
\vbox{
\hbox{
\epsfig{file=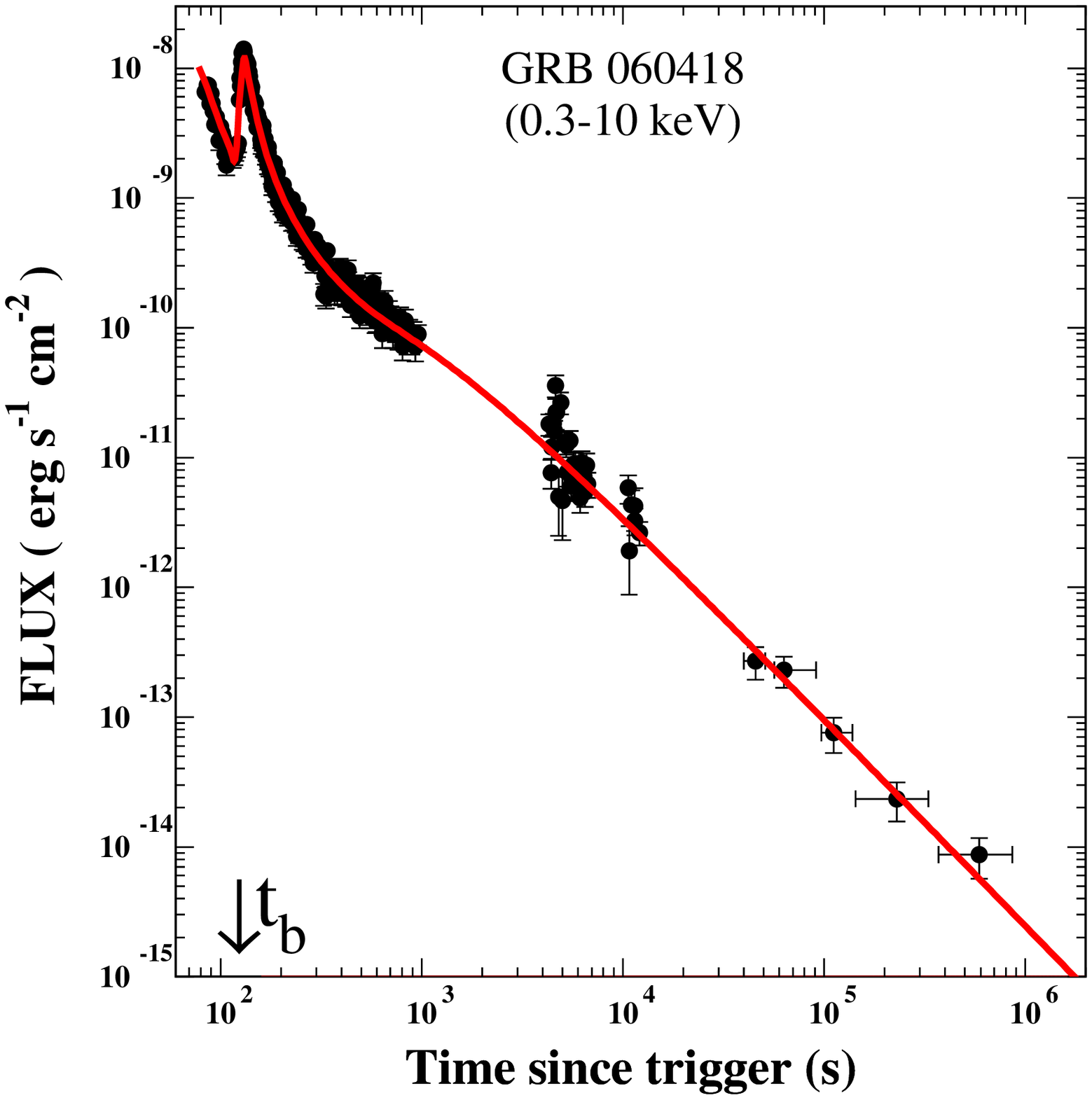,width=8cm} 
 \epsfig{file=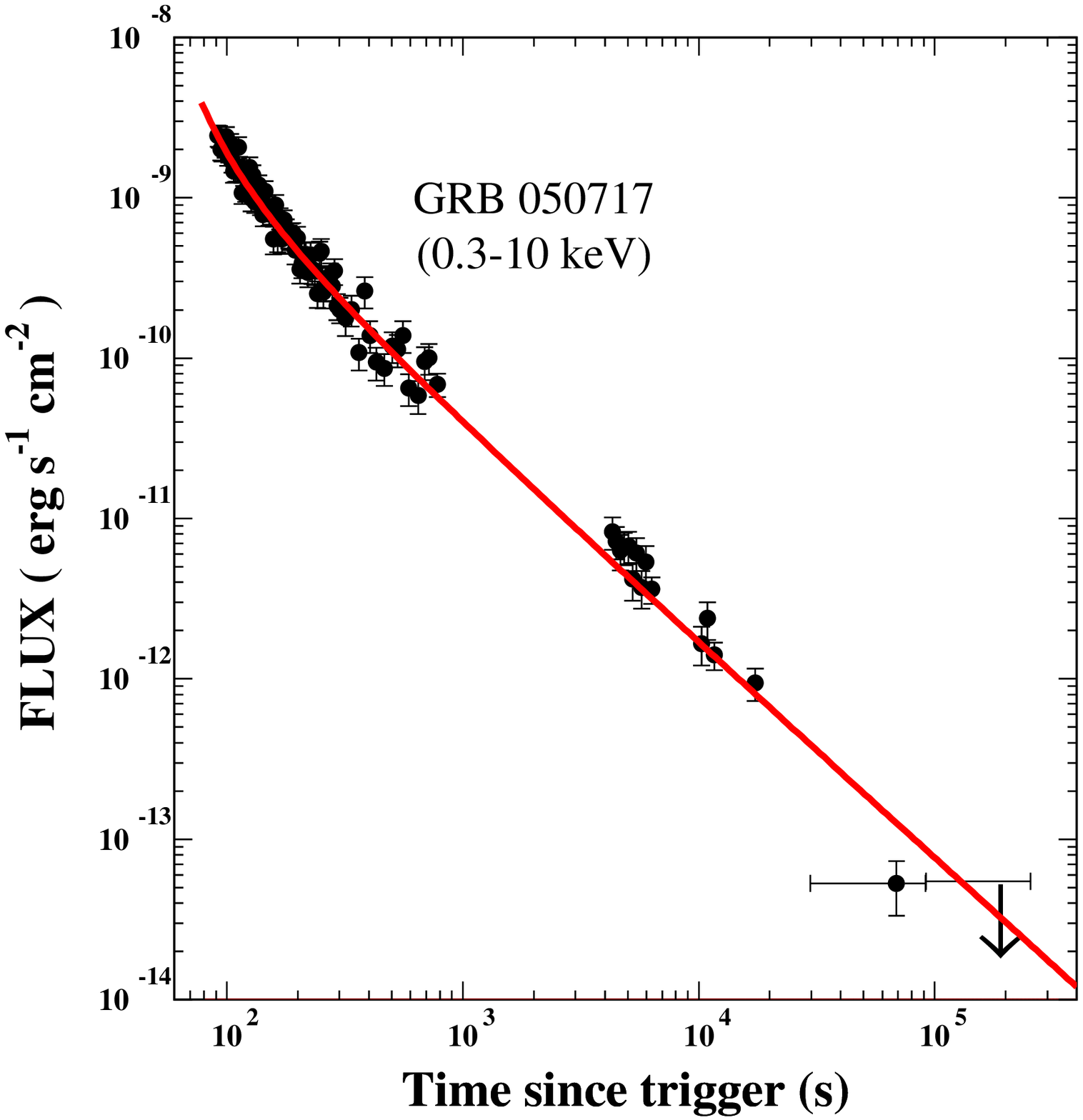,width=8cm} 
}}
\caption{
Comparison between the observed X-ray light curves 
of selected GRBs and their CB model fit: 
{\bf Top left (a):} GRB 070508.
{\bf Top right (b):} GRB 060813.
{\bf Bottom left (c):}  GRB 060418.
{\bf Bottom right (d):} GRB 050717.
The light-curve data
are  from the Swift/XRT light curve repository
(Evans et al.~2007). 
}
\label{f3}
\end{figure}

\begin{figure}[]
\centering
\vspace{-1cm}
\vbox{
\hbox{
\epsfig{file=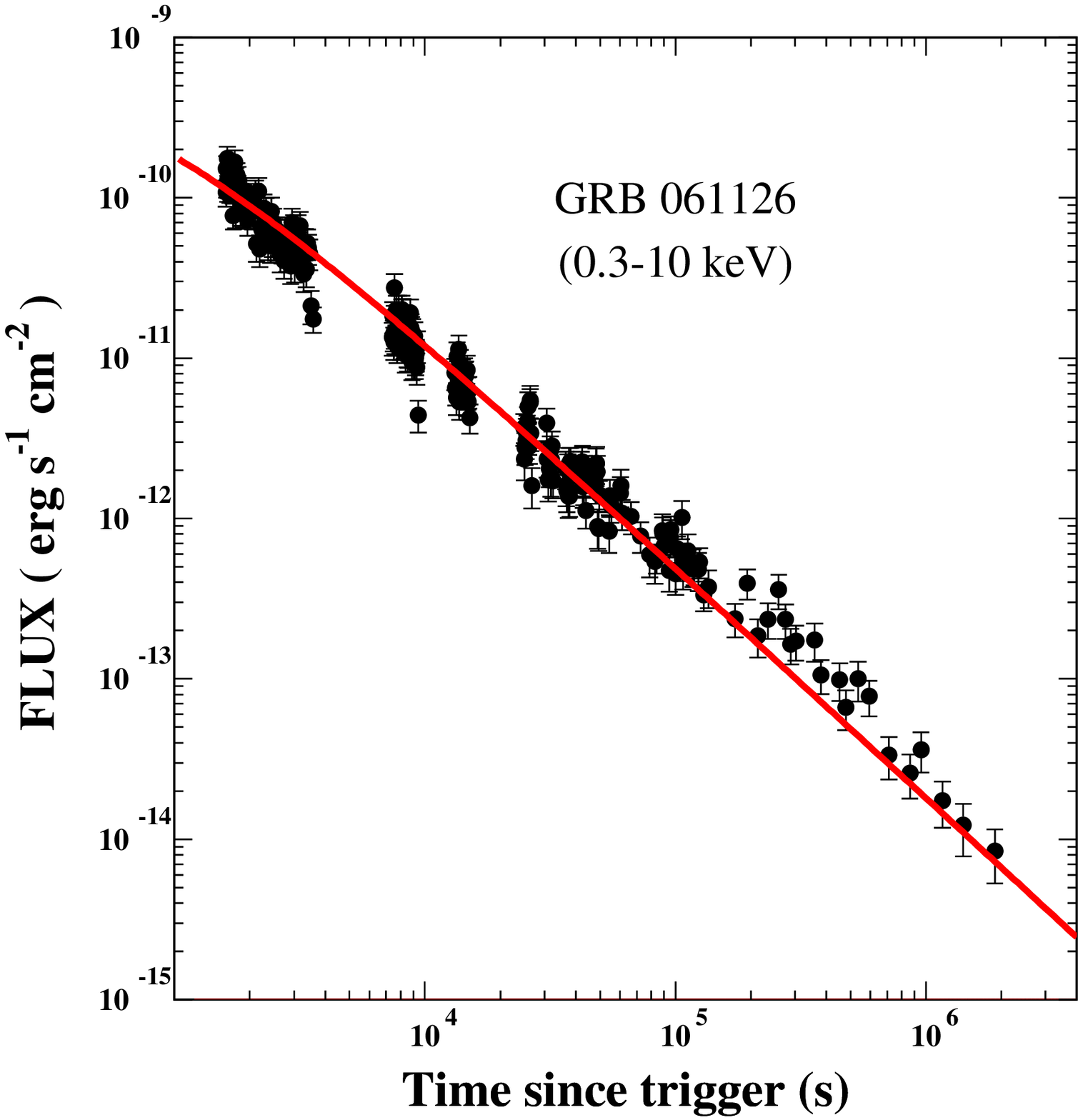,width=8.0cm} 
\epsfig{file=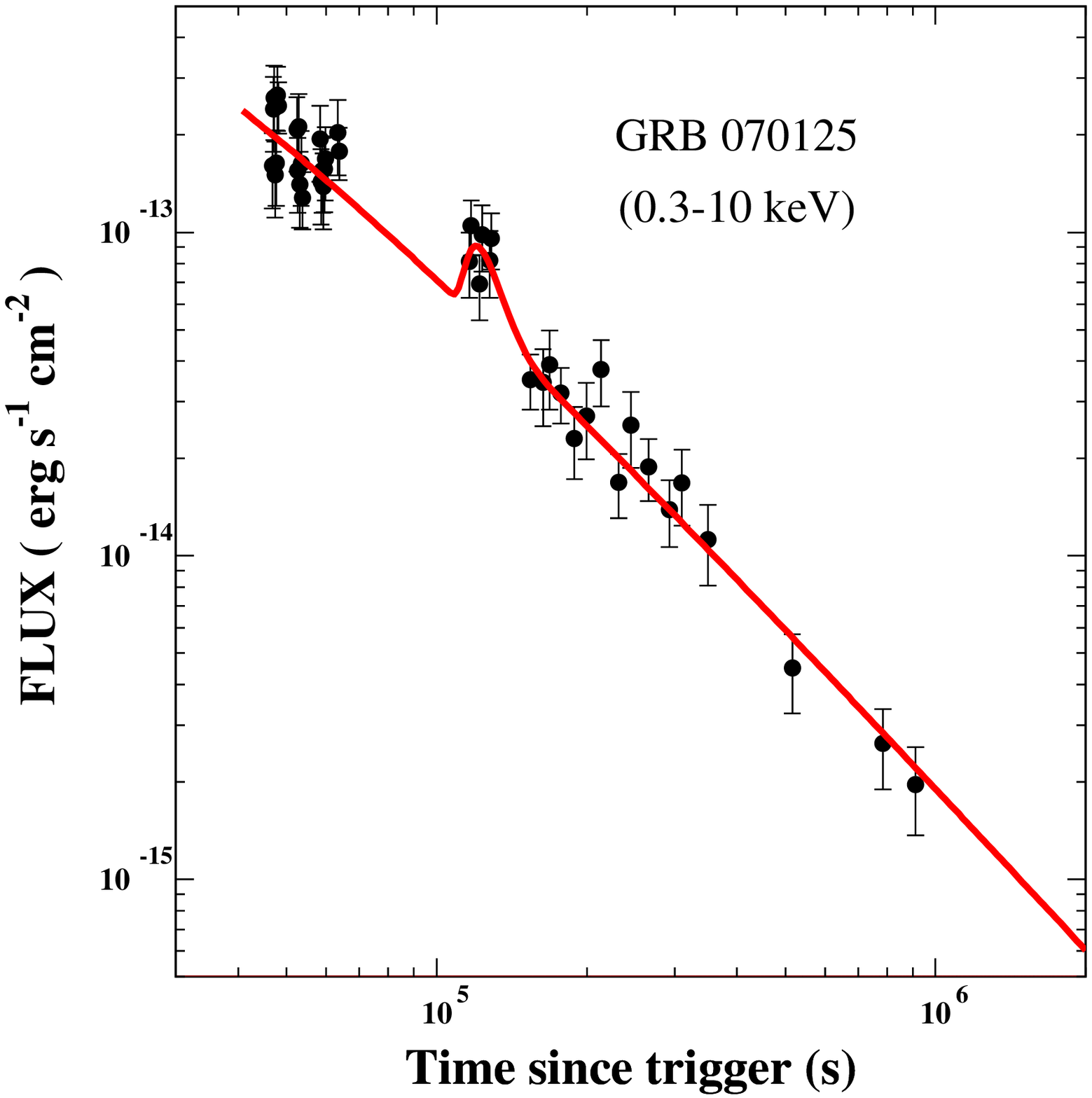,width=8.0cm} 
}}
\vbox{
\hbox{
\epsfig{file=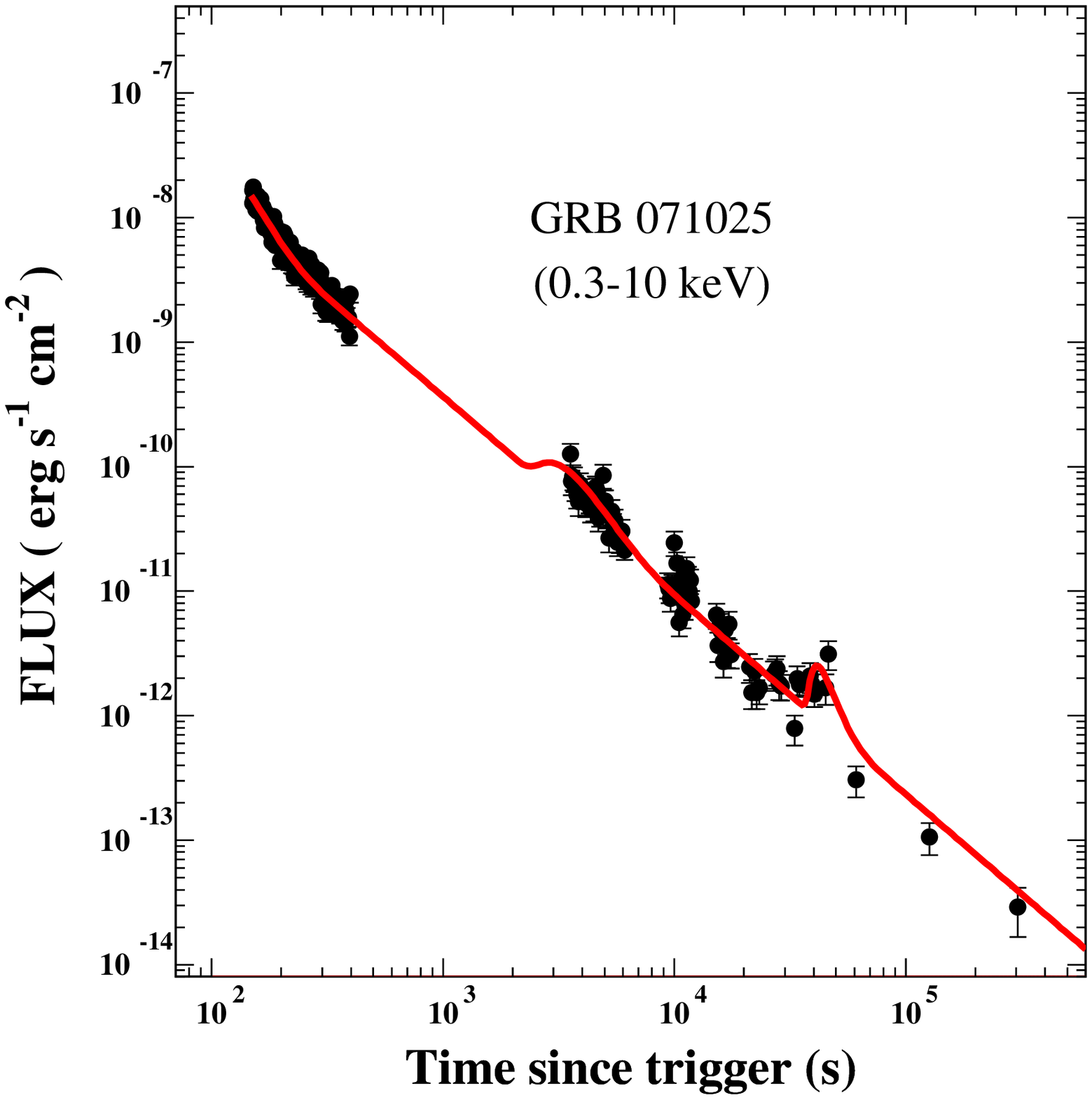,width=8cm} 
 \epsfig{file=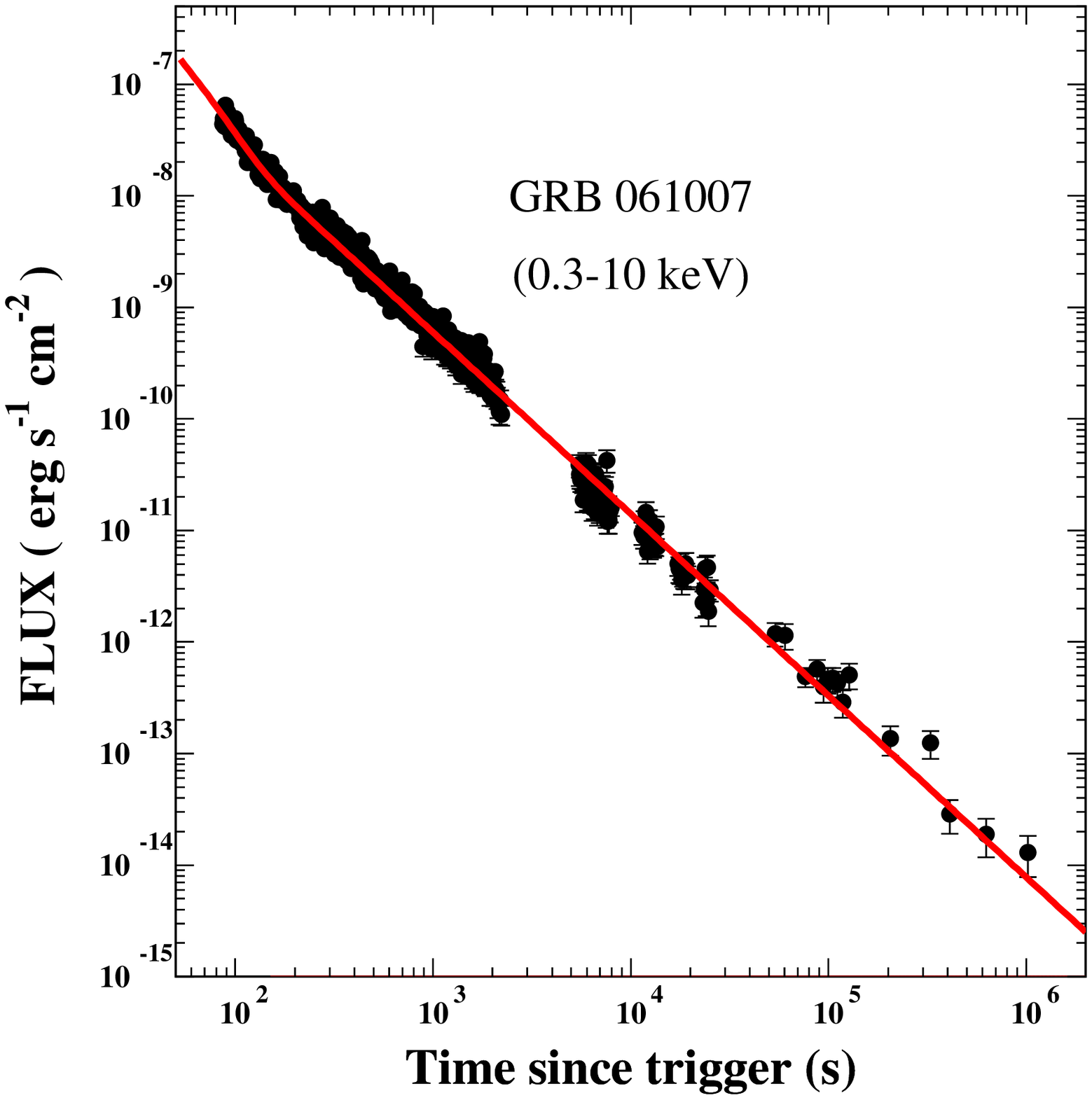,width=8cm} 
}}
\caption{
Comparison between the observed X-ray light curves 
of selected GRBs and their CB model fit: 
{\bf Top left (a):} GRB 061126.
{\bf Top right (b):} GRB 071025.
{\bf Bottom left (c):}  GRB 070125.
{\bf Bottom right (d):} GRB 061007.
The light-curve data
are  from the Swift/XRT light curve repository
(Evans et al.~2007). 
}
\label{f4}
\end{figure}

\begin{figure}[]
\centering
\vspace{-1cm}
\vbox{
\hbox{
\epsfig{file=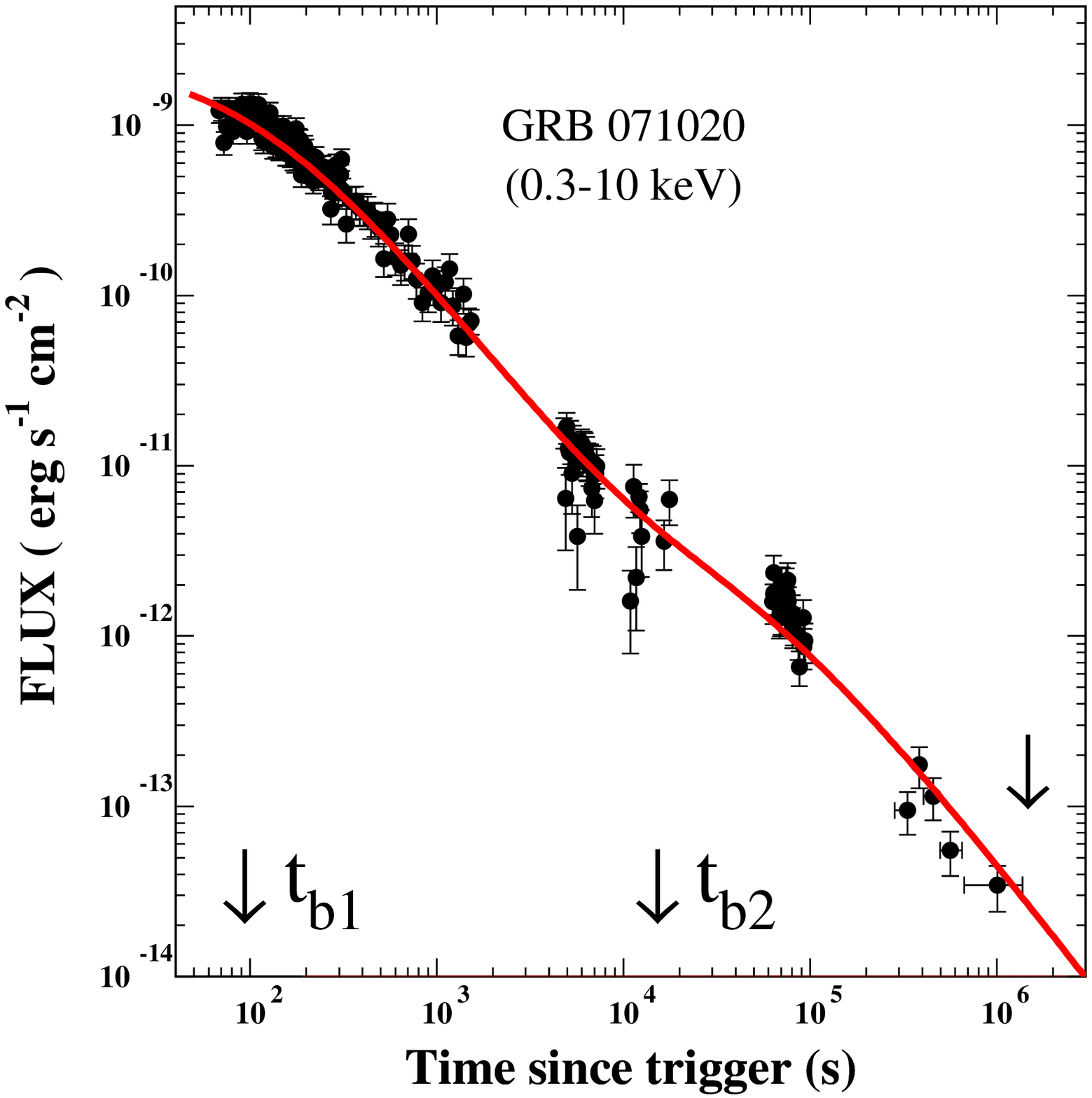,width=8.0cm} 
\epsfig{file=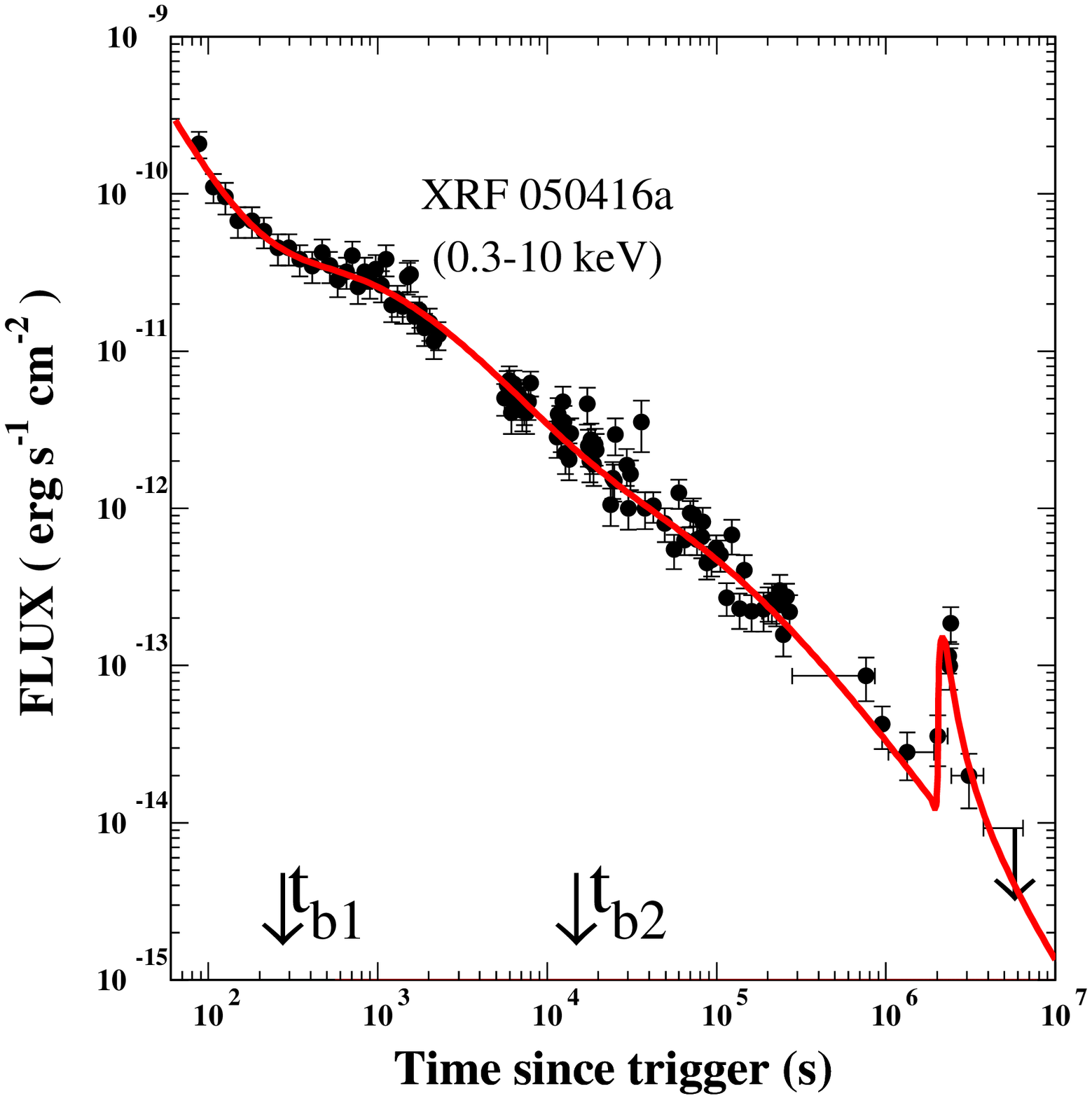,width=8.0cm} 
}}
\vbox{
\hbox{
\epsfig{file=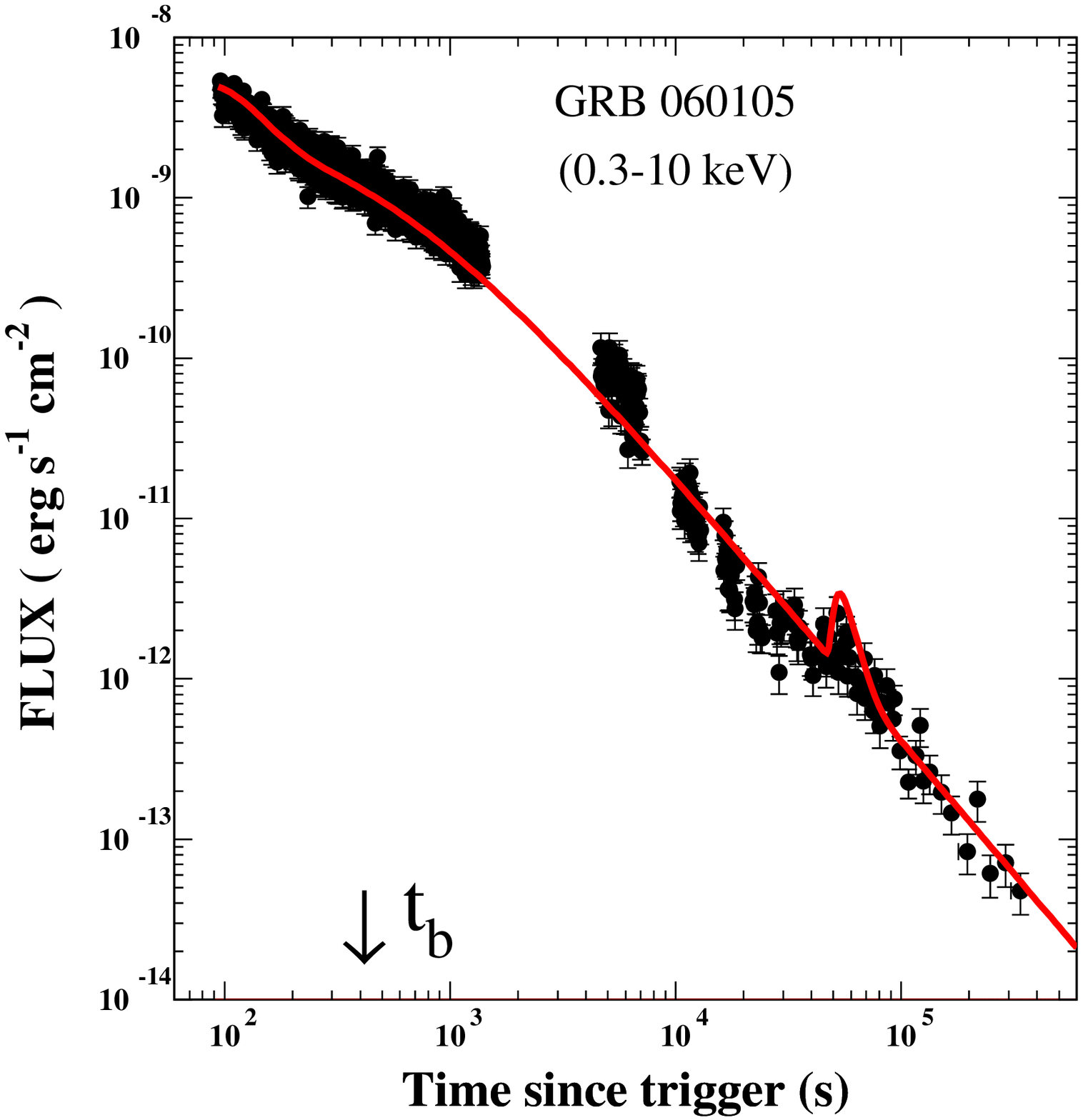,width=8cm} 
 \epsfig{file=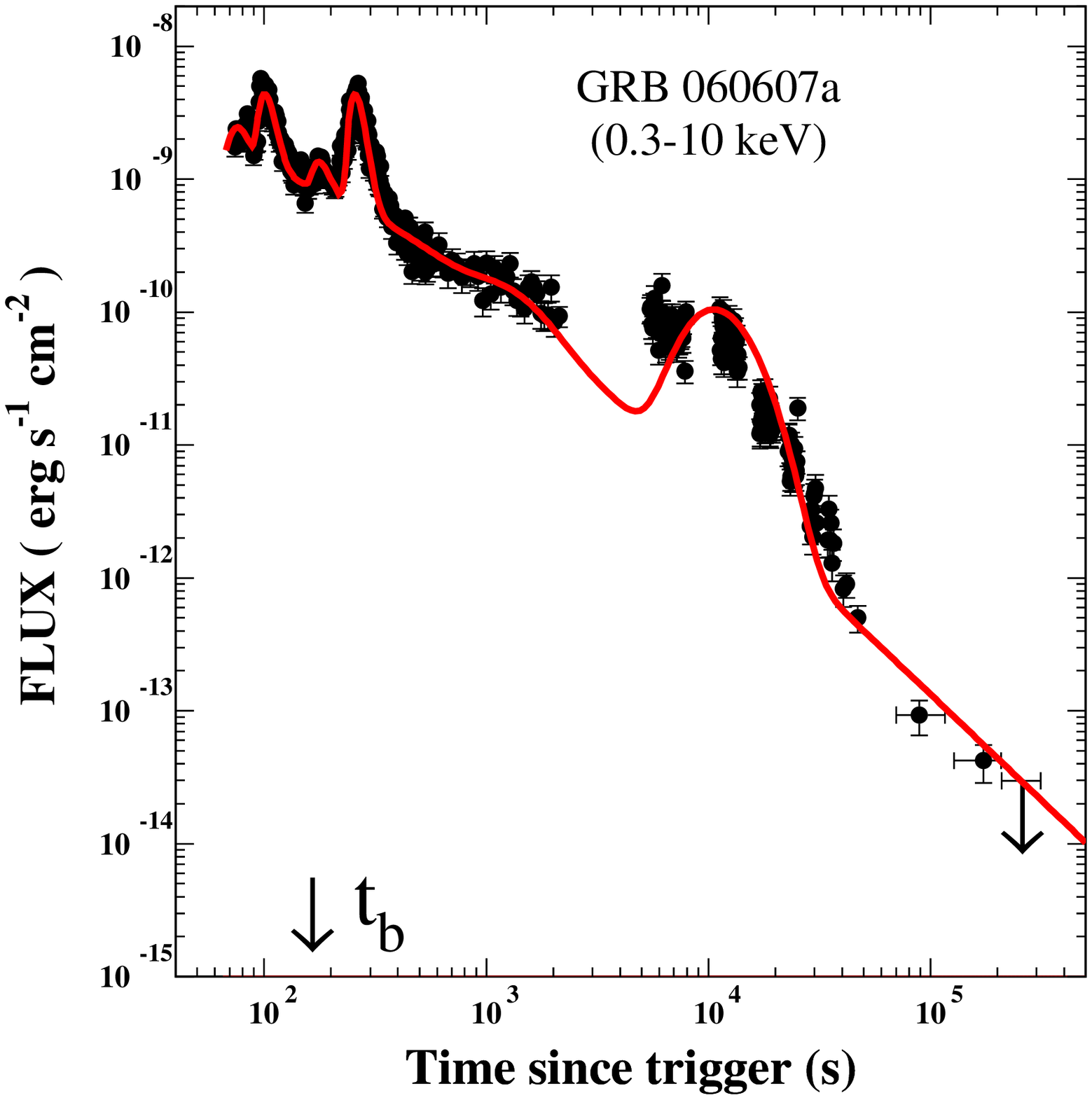,width=8cm}
}}
\caption{
Comparison between the observed X-ray light curves 
of selected GRBs and their CB model fit: 
{\bf Top left (a):} GRB 071020.
{\bf Top right (b):} GRB 050416A.
{\bf Bottom left (c):}  GRB 060105.
{\bf Bottom right (d):} GRB 060607A.
The light-curve data
are  from the Swift/XRT light curve repository
(Evans et al.~2007). 
}
\label{f5}
\end{figure}

In this section we comment one by one on the 16 GRBs or XRFs whose X-ray
light curves we discuss. The results of the CB-model fits
are shown in Figs.~\ref{f2} to \ref{f5}, and the parameters relevant to
our discussion are listed in Table \ref{t1}. The first eight GRBs are shown
in the order of decreasing $t_b$. For the next four, only an upper limit
on $t_b$ can be extracted from the fits. The last four have very complex AGs.

The presence or absence of visible breaks in X-ray light curves and their
different `look' --the panoply of possibilities that we illustrate with our
GRB choices-- depend not only on $t_b$ , but on its value relative to
$t_s$ (the start-time of XRT observations) and relative to the duration
of the initial period of prompt-radiation dominance over the synchrotron
AG. For this reason, it is easier to compare the plethora of looks of our GRBs 
in an order slightly different from that of a decreasing $t_b$.
This we do (only) in the next paragraph.

GRBs  980425 and 060729  have light curves with a complete
and simple canonical shape: one or two very clear prompt X-ray flares, a pronounced
fast decay, a long plateau, a very visible `break' smoothly bending at $t_b$ to
become a power-law decay. In GRBs 050401, 060105, 060418, 061007
and 050717, the plateau is becoming less and less pronounced, so that the AG's $t_b$
is hiding better and better under the prompt signal, to the point that the
last two are close to a pure power-law tail.  In GRBs 060813, 070508
and 050505, the prompt radiation ended early enough not to be caught by
Swift's XRT (in the last case the follow-up started very late), but this trio
displays very canonical AGs, with their neat plateaus softly bending into a late power law.
GRBs 071025, 061126 and 070125 are again approximate power-law tails,
in which neither the early X-ray flares nor the putative bend are seen.
 GRBs 071020, 050416A and certainly  060607A,
are very complex. The first two require contributions to the AG
from two distinct CBs, 050416A having also a late flare. The 
unsightly X-ray light
curve of GRB 060607A can be described by the CB model without any new 
ingredients, but not much is learned from fitting it.

\noindent 
{\bf GRB 980425.} The light curve of this 
memorable single-peak GRB, as observed by BeppoSAX (Pian et al.~2000),
is shown in Fig.~\ref{f2}a. The dotted line is the fit in DDD2002, 
showing what we called a pronounced `plateau'. 
We have added to it the last (predicted) data point, 
measured with Chandra by Kouveliotou et al.~(2004), some 1285 days after 
burst! To be consistent with the analysis here, we have re-fit the ensemble of 
data in the same manner as for all the other GRBs to be discussed.
The result is the continuous curve in the figure.
This GRB has, so far, the record large values
$\gamma_0\, \theta\!\approx\! 9.2$ and $t_b\!\approx\! 1.4\times 10^5$ s,
resulting in a light curve that rises before it falls, as explained in the
introduction and illustrated in Fig.~\ref{f1}a. This is the behaviour expected
for far-off-axis GRBs (DD2000a).
This one barely missed the official classification
as an XRF: it's $E_p$ is $\sim\!56$ keV, as opposed to $E_p\!<\! 50$ keV
(see Dado \& Dar~2005 for further comment on this point).

\noindent 
{\bf GRB 060729} and its X-ray light curve were studied in 
detail by Grupe et al.~(2007). It has a canonical shape, the longest 
follow-up observations with Swift XRT, and the record-high $t_b\!\sim\! 
8300$s, among the Swift GRBs. In Fig.~\ref{f2}b we show its CB-model 
description with, superimposed on its prompt decline phase, four ICS X-ray 
flares included in the fit, as discussed in detail in DDD2007a. This GRB 
being `canonical' and having a very clear `break' --as several others also 
discussed in DDD2007a-- is included here to illustrate the 
start of the transition from `breaks' to `missing breaks'.
Although the best fit to the X-ray AG appears to be excellent, it yields 
a large $\chi^2/dof\!=\!635/140\!=\!4.5$, mostly due to many
 far-flung isolated data points in the Swift data. More accurate data from XMM 
 Newton (Grupe et al.~2007) do not show such outliers. Eliminating their 
 contribution yields the $\chi^2$ value reported in Table 1.

\noindent
{\bf GRB 050505,} whose X-ray light curve was studied in
detail by Hurkett et al.~(2006). At $z\!=\!4.27$, this GRB is amongst the
most distant with a known redshift. Due to an Earth-limb
constraint, Swift was unable to slew to it until 47 minutes after the GRB's
trigger, and started measuring its X-ray light curve only 2883s after
burst, during the transition of the AG from its shallow-decline 
phase to a power-law decline.  As can be seen from Fig.~\ref{f2}c, 
the CB model describes very well 
the XRT light curve,
except when the counting rate becomes
comparable to the background.

\noindent 
{\bf GRB 050401,} whose X-ray light curve, studied in detail by De 
Pasquale et al.~(2006), evolves smoothly from the tail of the prompt 
emission at around 200s to a short decaying plateau, which suavely 
breaks into a power-law decline at $t\!\sim\!2000$s. Its CB-model description is
shown in Fig.~\ref{f2}d.
This GRB had a very bright X-ray AG,
even though it originated at a fairly large redshift, 
$z\!=\!2.9$, and had a very large 
extragalactic absorbing column density along its sight-line, inferred from 
its X-ray spectrum to be $N\!=\!1.7\times 10^{22}\, \rm cm^{-2}$ (De Pasquale 
et al.~2006), or $N\!=\! 4\times 10^{22}\,\rm cm^{-2}$ (Watson et 
al.~2006). Such a column density implies a very strong extinction of the
optical AG and, consequently, an extreme chromaticity:  more than 
10-magnitude extinction in the V band (Zombeck 1990) and more than 30 
magnitudes at 1800 $\rm \AA$. Indeed, the optical AG was very dim (a fitted 
spectral index, $- 0.33$, between the X-band and the optical band, 
compared to $ -0.8$ to $-1.1$ in `normal' GRBs.  In fact, according to 
Jakobson et al.~(2004), GRB 050401 qualified as a `dark burst'. It would not be 
a good case to discuss (unattenuated) chromaticity, or the lack 
thereof.

\noindent
{\bf GRB 070508.} Swift's XRT started to measure the X-ray light
curve of this GRB 82s after the GRB trigger. Even at this early time,
it  already displays the shallow-decay plateau phase of a canonical AG,
which later bends into a power-law decline, as shown in Fig.~\ref{f3}a.

\noindent 
{\bf GRB 060813}, shown in Fig.~\ref{f3}b, is a case in which the prompt 
radiation is not seen by the XRT, and the AG has no obvious flares. 
In spite of some evidence for local variability, the 
smoothly bending AG is well described  by the CB model
 ($\chi^2/${\it dof} $=\!1.07 $ for 254 {\it dof}). Had the break 
happened a bit earlier, as in other cases, the X-ray AG would look like a 
power-law.  The last data point lies below the fit, 
it could be due to an overestimated background.

\noindent
{\bf GRB 060418,} whose achromatic AG was studied in detail by Molinari et 
al.~(2007). Its X-ray AG evolves fast into a power-law decline, see 
Fig.~\ref{f3}c. The CB-model fit returns  
an early break at $t_b\!=\!123$s, well hidden under the flaring 
activity during the fast-decline phase of the prompt emission. 
The transition from an ICS-dominated 
regime to one in which SR is prevalent is corroborated by the fast 
spectral softening of the tail of the flare from around $t\!\sim\!130$s 
(Evans et al.~2007), which suddenly turns, at $t\!\sim\!165$s, into the 
much harder time-independent power-law spectrum characteristic of the 
synchrotron AG (DDD2007b). We have checked that the reasonable 
$\chi^2/${\it dof} $=\!1.21$ for 295 {\it dof} of the fit shown in the figure can be 
reduced to $\chi^2/${\it dof} $\sim\! 1$ by including X-ray flares between 5 
to 10 ks, or by replacing the fluctuating data points by average values.

\noindent 
{\bf GRB 050717,} studied in detail by Krimm et al.~(2006).
It had the largest inferred peak energy
of all Swift GRBs, $E_p\!=\!2401(- 568/+781)$ keV, despite its estimated large
redshift, $z\!>\!2.7\, .$ At this $z$-limit, 
$E_{\rm iso}\!\sim\! 1.1\times 10^{54}$ erg, and  the
local peak energy is $(1+z)\,E_p\!\sim\! 8840$ keV.
It also had 
an initially very bright X-ray AG, after the fast declining prompt
emission, with a power-law decline from $t\!\sim\! 200$s onwards.
The fit in Fig.~\ref{f3}d returns an early break-time limit,
$t_b\!<\!55$s, well hidden under the prompt-emission tail. 
The  CB-model interpretation of the transition from a prompt ICS
radiation to a synchrotron AG is supported by
the observed rapid spectral softening of
the tail of the prompt emission and its sudden change at $t\!\sim\! 200$s
into the harder time-independent power-law spectrum of the
synchrotron AG (DDD2007b). In the case of this GRB 
the best-fit value, $p=1.67$, does not
satisfy Eq.~(\ref{Asymptotic}), with $\Gamma=1.61\pm 0.10$,
as inferred from the 
X-ray spectrum with a fixed column density limited to the Galactic one
(Krimm et al.~2006).  However, $p=1.67$ is consistent with
$\Gamma=1.88\pm 0.15$ of the AG for $t>200$s, the spectral index 
reported by Zhang et al.~(2007),
after inclusion of host-galaxy and IGM absorption.

\noindent
{\bf GRB 061126,}   
studied in detail by Perley et al.~(2008) had two major prompt pulses. 
Due to an Earth-limb constraint, Swift 
slewed to the burst's direction only  23 minutes after its
localization by its Burst Alert Telescope (BAT). Its light curve,
measured by the XRT between 1.6 ks and 1.88 Ms, is shown in Fig.~\ref{f4}a.
The X-ray light curve was reported to be  well fit  by a  power-law in 
time with index  $1.29\pm 0.08$ (Sbarufatti et al.~2006).
A CB-model fit, with  $p\!=\!1.89$ and $t_b<104$s, is shown in 
Fig.~\ref{f4}a. There is a possible indication in the data 
of a steeper decay between 1.6~ks and 3.6~ks, which might
belong to the tail of another CB with a smaller $t_b$.
Cases of AGs clearly requiring two CBs will be discussed anon.

\noindent
{\bf GRB 071025.} Swift's XRT  started observations 
of the X-ray light curve 146s after the BAT trigger. The initial 
relatively hard spectrum ($\Gamma\!=\!1.4$) softened beyond 300s 
and the light curve declined like a single power-law, consistent with 
the CB-model's asymptotic power-law decline with a power-law index 
$\simeq\!1.6$, as shown in Fig.~\ref{f4}b. The data suggest a flaring 
activity 
between 4 ks and 40 ks. The effect of such flares on the CB-model 
X-ray light curve is illustrated in the figure by adding an ICS flare 
around 40 ks with parameters (peak time, width and normalization)
chosen, as in all other cases with clear flares, to best fit the data.

\noindent 
{\bf GRB 070125,}   studied in detail by Bellm et al.~(2007).
It was detected by Mars Odyssey, Suzaku, INTEGRAL, and RHESSI.
It is one of the Swift-era GRBs with the largest
measured values of $E_{\rm iso}\!\sim\! 1\times 10^{54}$ erg, 
$L_p\,\sim\, 3\times 10^{53}\, {\rm erg\, s^{-1}}$, and
source-frame $(1+z)\, E_p\, \sim\, 1100$ keV.
The initial detection of this GRB occurred while it was not in the BAT 
field of view during the beginning of the prompt emission,
and its XRT light curve starts at 46 ks after the burst.
 As shown in Fig.~\ref{f4}c
its power-law decline is well described by the CB model.
The feature at $\sim \! 110$~ks can be interpreted as an X-ray flare,
as in the figure.

\noindent
{\bf GRB 061007,} 
whose AG was studied in detail by Schady et al.~(2006) and Mundell et al.~(2007), 
was  the brightest GRB detected by Swift  and was 
accompanied by an exceptionally luminous X-ray and UV/optical
afterglow, which decayed as a power law with an index $1.65\pm 0.02$.
It had the largest values of 
 $E_{\rm iso}\!\sim\! 1\times 10^{54}$ erg, 
 $L_p\!\sim\! 2\times 10^{53}\, {\rm erg\, s^{-1}}$
 and an emission-point
peak energy, $(1+z)\, E_p \!\sim\! 1000$ keV
(Golenetskii et al.~2006).
This GRB is the best example to date of a bright
X-ray AG,  well-sampled from the start of the XRT observations
(86s after the BAT trigger) to $10^6$ s.
The AG, shown in Fig.~\ref{f4}d, is power-law behaved right after the 
tail of the prompt emission. The CB-model fit returns $t_b\!<\!89$s, below 
which the $\chi^2/${\it dof} (a reasonable 1.13 for 1030 {\it dof)} stays put.

\noindent
{\bf GRB 071020,}  measured by Swift's XRT
between 68s and 1.7 Ms after trigger, and shown in Fig.~\ref{f5}b.
Holland et al.~(2007) fitted the data with a broken power-law with 
an initial decay index of $\approx \! 0.5$, a break at
$t_b\,=\,160$s, and a late-time decay index of
$1.14 \pm 0.02$. 
The fit is poor between 1.5 ks and 1.5 Ms. 
A CB-model fit with a single CB is also unsatisfactory.  
The addition of a second CB to the AG's description, as in the fit
shown in Fig.~\ref{f5}a, greatly improves the fit to
$\chi^2/${\it dof} $=\!1.52$ for 174 {\it dof}, acceptable in view of
what appears to be evidence for  
flaring activity, from 1.5 to 15 ks, which we have not
endeavoured to describe, given the scarcity of data.

\noindent
{\bf XRF 050416A.}
The complex X-ray light curve of this XRF was monitored up to 74 days after 
the burst (Mangano et al.~2007).
  The late decline rate of the light curve is
significantly slower than expected in the CB model from the observed
photon spectral index $\Gamma$, namely $t^{-\Gamma-1/2}\!\sim\!
t^{-1.5\pm 0.10}$.  The prompt signal of XRF 050416A had  two clear
 pulses which, in the CB model, correspond to two separate CBs.
The X-ray light curve, modeled with two CBs and shown
in Fig.~\ref{f5}b, has a SR-component late-power decay that
--although it is not readable `by eye' due to the late-occurring
ICS flare-- is compatible with the
predicted one.

\noindent 
{\bf GRB 060105,} whose X-ray light curve was studied in detail 
by Tashiro et al.~(2007). Following the prompt emission, which ended with 
a very steep decay, the light curve is canonical, it has a shallow decay 
after 180s and steepens at around 500s to a fast power-law decline, with a 
weak flaring activity superposed on it. 
The deviations from a smooth X-ray light curve may be caused by the flaring 
activity, not included in this particular fit, whose 
$\chi^2/${\it dof} $=\! 1.36$ for 854 {\it dof,} is not inadequate.

\noindent
{\bf GRB 060607A,} was studied by  Molinari et 
al.~(2007). Its complex X-ray light curve, like that of quite a few other
GRBs, is dominated by strong flaring activity, as can be seen
in Fig.~\ref{f5}d,
with its many flares superimposed on the AG of a fitted, single, dominant CB.
This fit, which can be improved by splitting the
last flare into two, is a very rough description
($\chi^2/${\it dof} $=\! 4.9$ for 440 {\it dof}),
not a proof of the quality of a prediction.
Moreover, in cases with such a prominent flaring activity,
the photon spectral index of the AG data  
is an average between the typical index of flares,
$\Gamma=1$, and that of synchrotron AG, $\Gamma=2$,
i.e., an average significantly smaller than that of the synchrotron AG. 
Thus, we do not expect such a labyrinthine AG to satisfy 
the CB-model spectral-index relations, Eqs.~(\ref{Fnux},\ref{Asymptotic}).

\subsection*{The afterglow as a function of time and frequency}

We have summarized in Eq.~(\ref{Fnu}) the predicted form of the spectral
energy density of the AG of a GRB, in which the
time-dependence and the energy-dependence are explicitly concatenated. In
the large-frequency limit of the X-ray domain, the expression simplifies
to that of Eq.~(\ref{Fnux}), implying a predicted relation between the temporal 
index $p$ (which we fit to the XRT light curve of the X-ray AG)  and the
spectral index $\Gamma$, independently fitted by the Swift team to the X-ray 
AG spectrum after correcting for attenuation, 
and reported by Zhang et al.~(2007). 
The prediction is particularly simple, and
is most transparently readable in the late-$t$ limit for the AGs'
dependence on $t$ and $\nu$, Eq.~(\ref{Asymptotic}), in which both the
time and the frequency functional forms are separate power laws.

The values of $p$ and $\Gamma$ are listed in Table \ref{t1}.  Notice that
$\Gamma$ varies over a significant range of central values, 1.61 to 2.25,
and that the measurements are  not compatible within errors with
a common value. To illustrate the prediction in Eqs.~(\ref{Fnux},\ref{Asymptotic}), we
have plotted in Fig.~\ref{f6} the ratio $r\!=\!p/(2\, \Gamma -2)$
(predicted to be unity) for the various GRBs analized in this paper,
and added a few other analized in the same fashion.
The results are quite
satisfactory. The mean value of $r$, for instance, is $0.999\pm 0.025$
for the GRBs analized here, $1.000\pm 0.019$ for the ensemble
plotted in Fig.~\ref{f6}.

\subsection*{The $(t_b,\,E_{\rm iso})$ and $(t_b,\,E_p)$ correlations}

In the CB model, the functional dependence on $\theta$ and $\gamma_0$ of 
the deceleration-bend time  of the synchrotron AG, $t_b$, 
as well as its normalization, are specified by Eq.~(\ref{tbreaks}). 
This is also the case for the  parameters, $E_p$ and $E_{\rm iso}$
of Eqs.~(\ref{ep}) and (\ref{eiso}), of the prompt ICS signal.
As we saw in the introduction, this implies explicit correlations 
between $t_b$ and the prompt observables. The $(t_b,\,E_{\rm iso})$
correlation
is illustrated in Fig.~\ref{f1}b for various choices of $\theta$ and $\gamma_0$,
with the rest of the parameters in $t_b$ and $E_{\rm iso}$ fixed to
 reference values in Eqs.~(\ref{tbreaks},\ref{eiso}).

In Fig.~\ref{f7} we  plot, in the $[t_b/(1+z),\,E_{\rm iso}]$ plane,
the values returned by our analysis of
the GRBs we have discussed, see Table \ref{t1}.
The GRBs represented by arrows reflect the fact that some
data are just upper limits.
The large shaded contour plot in the figure is the boundary of the
domain covered by letting $\gamma_0$ vary from 500 to 1500,
$\theta$ from 0 to 8 mrad, typical ranges encountered
in the CB-model analysis of GRBs. Moreover, the normalization 
of $t_b$ in Eq.~(\ref{tbreaks}) was varied from its central value 
in Eq.~(\ref{tbreaks}) to 1/2 order of magnitude above it, and
the normalization of $E_{\rm iso}$ in Eq.~(\ref{eiso})
from its central value to 1/2 order of magnitude below and above
it. The variability in these normalizations is best
ascertained by the current analysis, it has been chosen to make
Fig.~\ref{f7} `look good'. We have added to the figure the
results for a few GRBs which we have previously analyzed
along the same lines in DDD2007a,b.

There is no reason to expect the data to populate uniformly
the region bounded by the contour in Fig.~\ref{f7}. On the
contrary. The relativistically beamed radiation from a 
point in a CB initially subtends an angle $1/\gamma_0$.
Observers at an angle $\theta$ from the axial direction have 
a chance $\propto\! \theta\,d\theta$ of being illuminated. 
At $\theta\! > \! 1/\gamma_0$ this chance decreases abruptly,
given the fast fall of the Doppler factor. All in all, 
$\theta\!\sim\!1/\gamma_0$ is the optimal observation angle,
for {\it any} $\gamma_0$. Most GRBs, then, should be seen
at $\theta\,\gamma_0\!=\! {\cal{O}}(1)$. The thick straight line
in Fig.~\ref{f7} is $t_b(E_{\rm iso})$ at fixed $\theta\,\gamma_0$,
for which $t_b/(1+z)\!\propto\!\gamma_0^{-3}$ and  
$E_{\rm iso}\!\propto\! \delta_0^3\!\propto\!\gamma_0^{3}$. Thus:
\begin{equation}
t_b/(1+z) \propto E_{\rm iso}^{-1}.
\label{naivetbEiso}
\end{equation}
The data follow this trend well, but at the
high-$E_{\rm iso}$ end, at which they bend as in Fig.~\ref{f1}b.

In Fig.~\ref{f8} we plot, in the $[t_b/(1+z),\,(1+z)\,E_{\rm p}]$ plane,
the corresponding results of our analysis. The shaded domain is obtained
with the same ranges in $\gamma_0$, $\theta$ --and in the normalization
of $t_b$-- as in the previous paragraph. The normalization of $E_p$ has
been allowed to vary from 1/3 to 1/6 of its value in
Eq.~(\ref{ep})\footnote{The $E_p$ of Eq.~(\ref{ep})
is the peak energy at the start of a  pulse; set it to $t\!=\!0$. The energy of
the radiation is predicted to decrease during the pulse's duration:
$E_p(t)\!\approx\!E_p(0)\,[1-t/(\Delta^2+t^2)^{1/2}]$, with  $\Delta$ the width 
parameter (the full width at 
half-maximum, FWHM, is $\sim\!1.8\,\Delta$). Observers usually report 
$E_p$ at the peak's maximum,
expected to be $E_p(t_{\rm max})\!\approx\! 0.23\,  E_p(0)$, or
its pulse-averaged value: $\langle E_p\rangle\!\approx\! 0.18\,  E_p(0)$
over the FWHM. We have not 
corrected for these facts, which may explain the choice of the `best' domain.}.
The points plotted as ellipses have an
unknown $z$, which we have let vary from 0 to 2.75, the average
for Swift-era GBRs (Greiner, $http://www.mpe.mpg.de/~jcg/grbgen.html$).  
At fixed $\theta\,\gamma_0$, 
$(1+z)E_p\!\propto\!\gamma_0\,\delta_0\!\propto\!\gamma_0^2$, so that:
\begin{equation}
t_b/(1+z) \propto [(1+z)\,E_p]^{-3/2}.
\label{naivetbEp}
\end{equation}
The rest of the comments are
as in the  discussion of the $(t_b,\,E_{\rm iso})$ correlation.

\begin{figure}[]
\centering
\vspace{-1cm}
 \epsfig{file=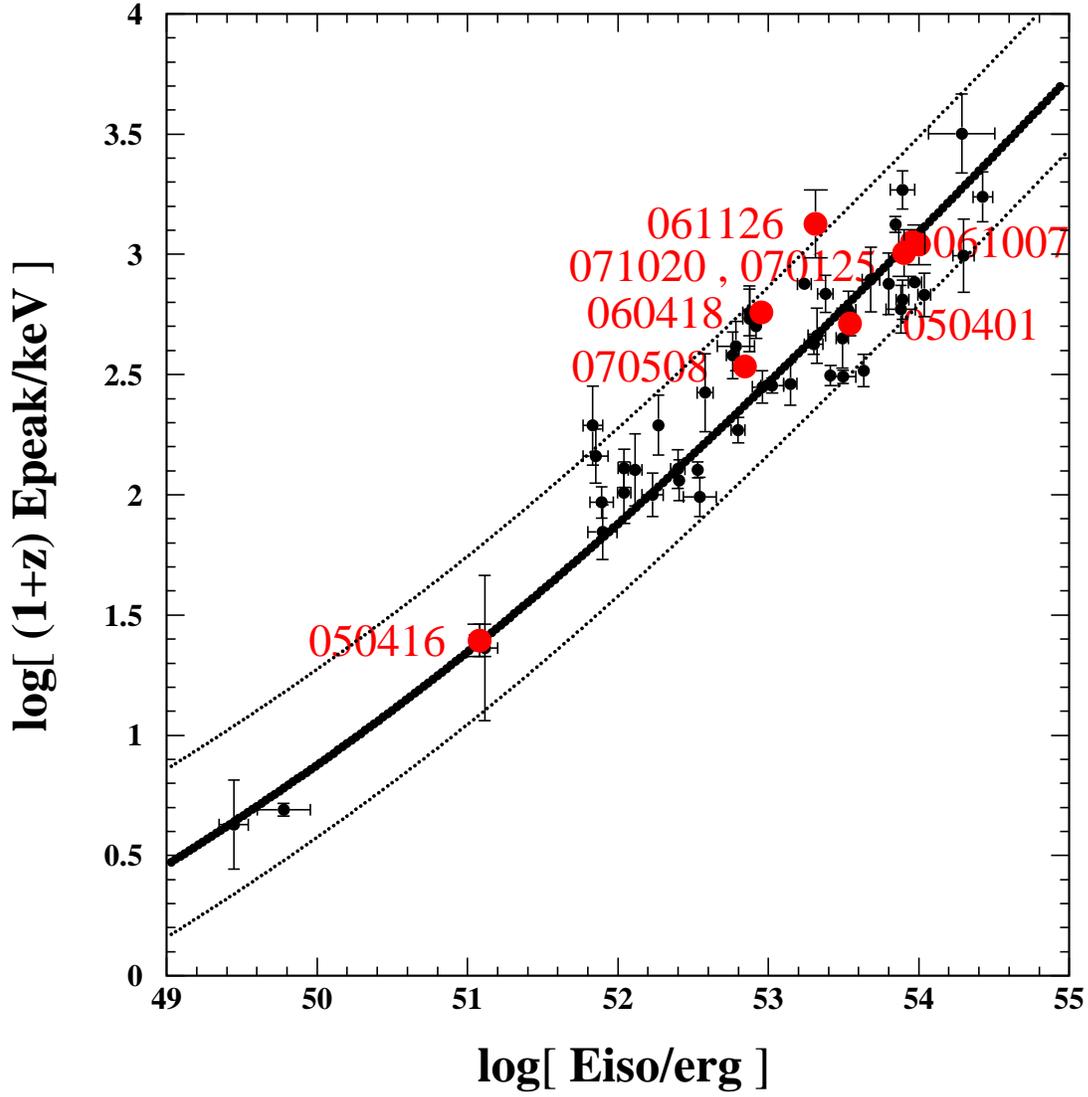,width=16cm}
\caption{
The $[(1+z)\,E_p,\,E_{\rm iso}]$ correlation (DD2000b,
Amati et al.~2002) for an ensemble of GRBs of known $z$, analized by Schaefer (2007).
The central line is the CB-model's expectation (DDD2007c), the dotted lines
bracket the observed case-by-case variability. The large (red) circles are the
GRBs discussed in this paper which have known $z$, $E_p$ and $E_{\rm iso}$.
}
\label{f9}
\end{figure}

Another direct way to ascertain the variability of the parameters
governing the normalizations of $E_p$ and $E_{\rm iso}$ is to study their
scatter plot (DD2000b, Amati et al.~2002, DD2004, 
DDD2007c, Amati 2006)  for a large collection of GRBs and XRFs.
This is done in Fig.~\ref{f9}, where the varying-power correlation
predicted by the CB model (DDD2007c) is shown, and to which the GRBs with
known $z$, $E_p$ and $E_{\rm iso}$, among those studied here, are
added. The figure shows that a total uncertainty of a factor of 2 in the norm
of $E_p$ and of one an order of magnitude in the norm of $E_{\rm iso}$ (as
we have adopted) is adequate to bracket the data.

We have also tested elsewhere (DDD2007d) the correlation, apparent
in Fig.~\ref{f1}a, between $t_b$ and the normalization of the AG.
Willingale et al.~(2007) and Nava et al.~(2006) had collected and analyzed 
a large set
of GRBs, and made a scatter plot of $t_b$ versus the total AG energy in
the 15-150 keV X-ray band up to time $t_b$. To use this available information,
we studied this correlation in its Willingale-Nava form. Like the ones
in Figs.~\ref{f7} and \ref{f8}, it turns out to follow the pattern expected in the CB model.

The correlations between $t_b$ and $E_{\rm iso}$ or $E_p$ demonstrate that
`sub-energetic' GRBs (or XRFs) have large `break' times and, consequently,
easily observable `breaks'. As GRBs become more `energetic', $t_b$
decreases and the chances increase to `miss the break', which may be
hidden under the prompt radiation, or may precede the Swift slew-time
minimum, or the start of the observations.

\section{Conclusions and outlook}

A virtue of astrophysical X-ray data is that, in many instances and
relative to lower-frequency bands, the corrections for attenuation are
simpler and more reliable. The strength of Swift in dealing with transient
phenomena is, as the satellite's name reflects, the prompt start of its
data-taking after an alert. This has made the Swift results an excellent
testing ground for theories of GRBs and XRFs. In particular, the ability
to monitor the X-ray flux over a very wide range of times has provided
decisive tests of the theoretical predictions.

Filling the pre-Swift gap in the data --between `prompt' and `afterglow' radiations--
has led to a better understanding of the mechanisms responsible for
them. In the CB model they are different: inverse Compton scattering
 and synchrotron radiation, 
respectively. We have previously argued that the strong case for an ICS origin of the 
`prompt' radiation (DD2004) has been reconfirmed by the analysis of the Swift X-ray flares, 
and the fast decay of the prompt signals (DDD2007b).  In this respect 
the study of X-ray and optical data is also particularly meaningful 
(DDD2007a). 
The observed correlations between 
prompt observables ---$E_{\rm iso}$, $E_p$, $L_p$ and
pulse rise-time, lag-time and variability--- also agree with the CB-model
(see DDD2007c and references therein). These correlations follow
from the same simple considerations, that we have emphasized in this paper,
on the dependence of the cited prompt observables
on the Lorentz and Doppler factors of the radiation emitted by a quasi-point-like 
relativistically moving source.

\begin{figure}[]
\centering
\vspace{-1cm}
\vbox{
\hbox{
\epsfig{file=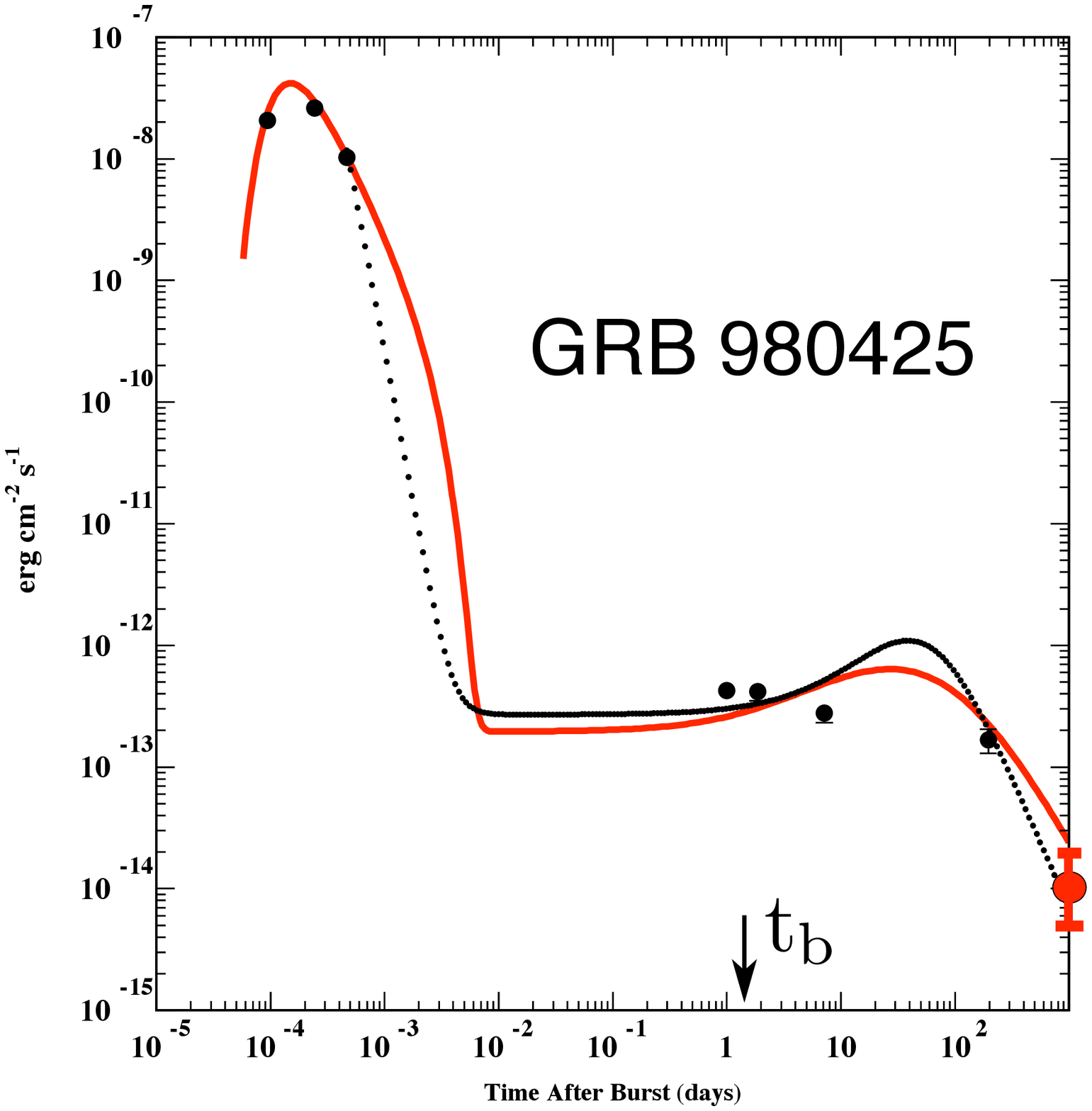,width=8.0cm}
\epsfig{file=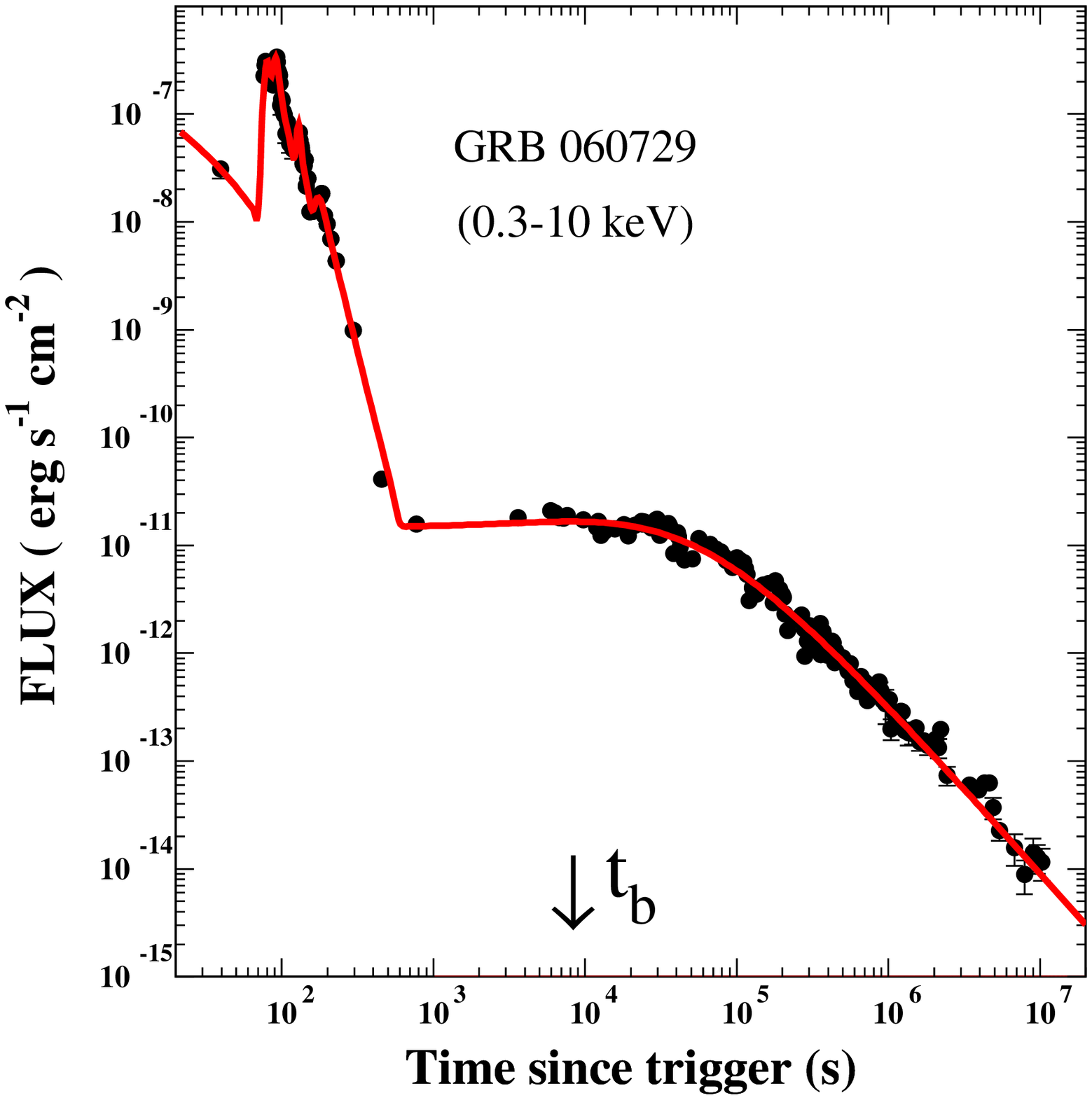,width=8.0cm} 
}}
\vbox{
\hbox{
\epsfig{file=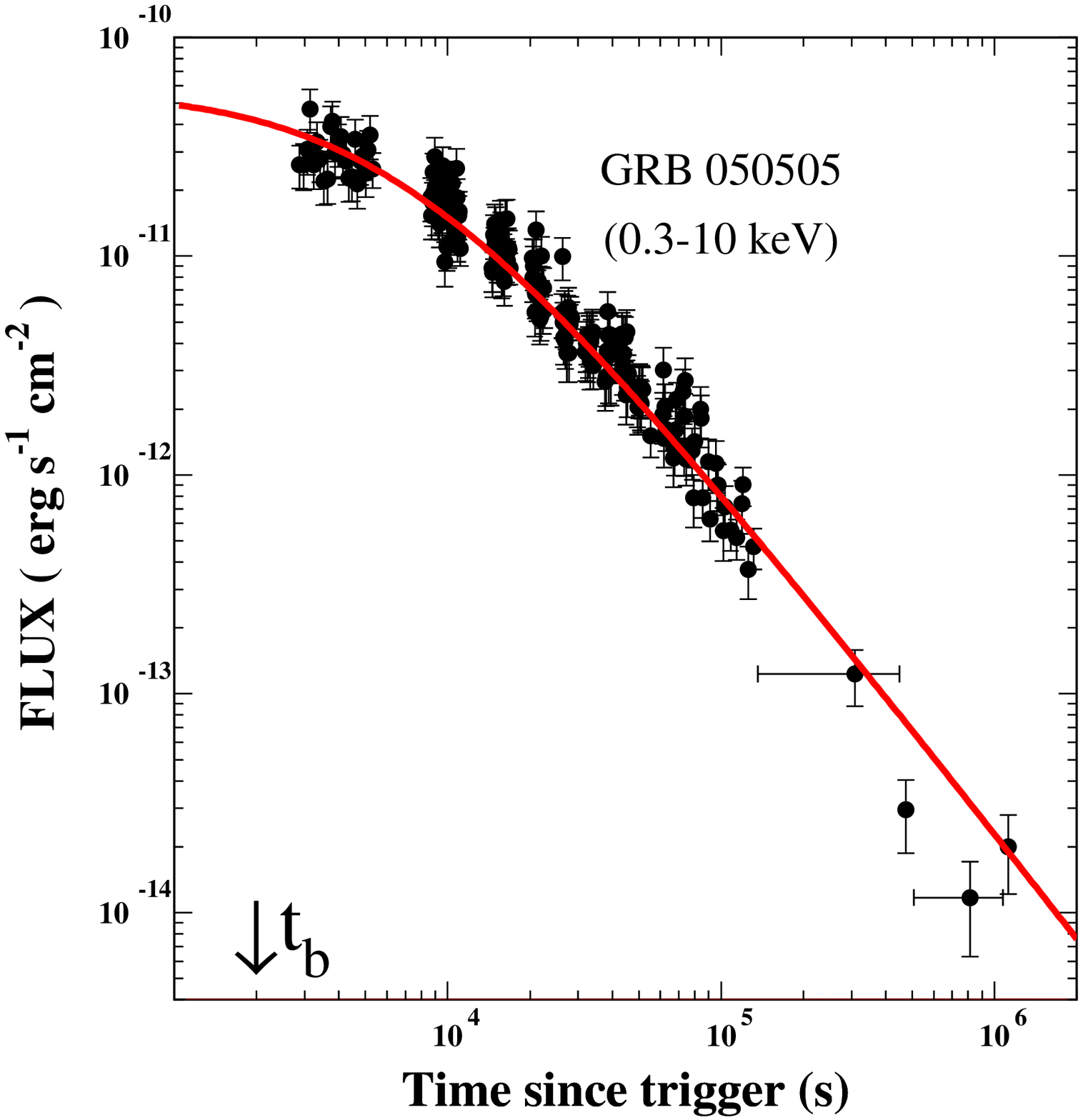,width=8cm} 
 \epsfig{file=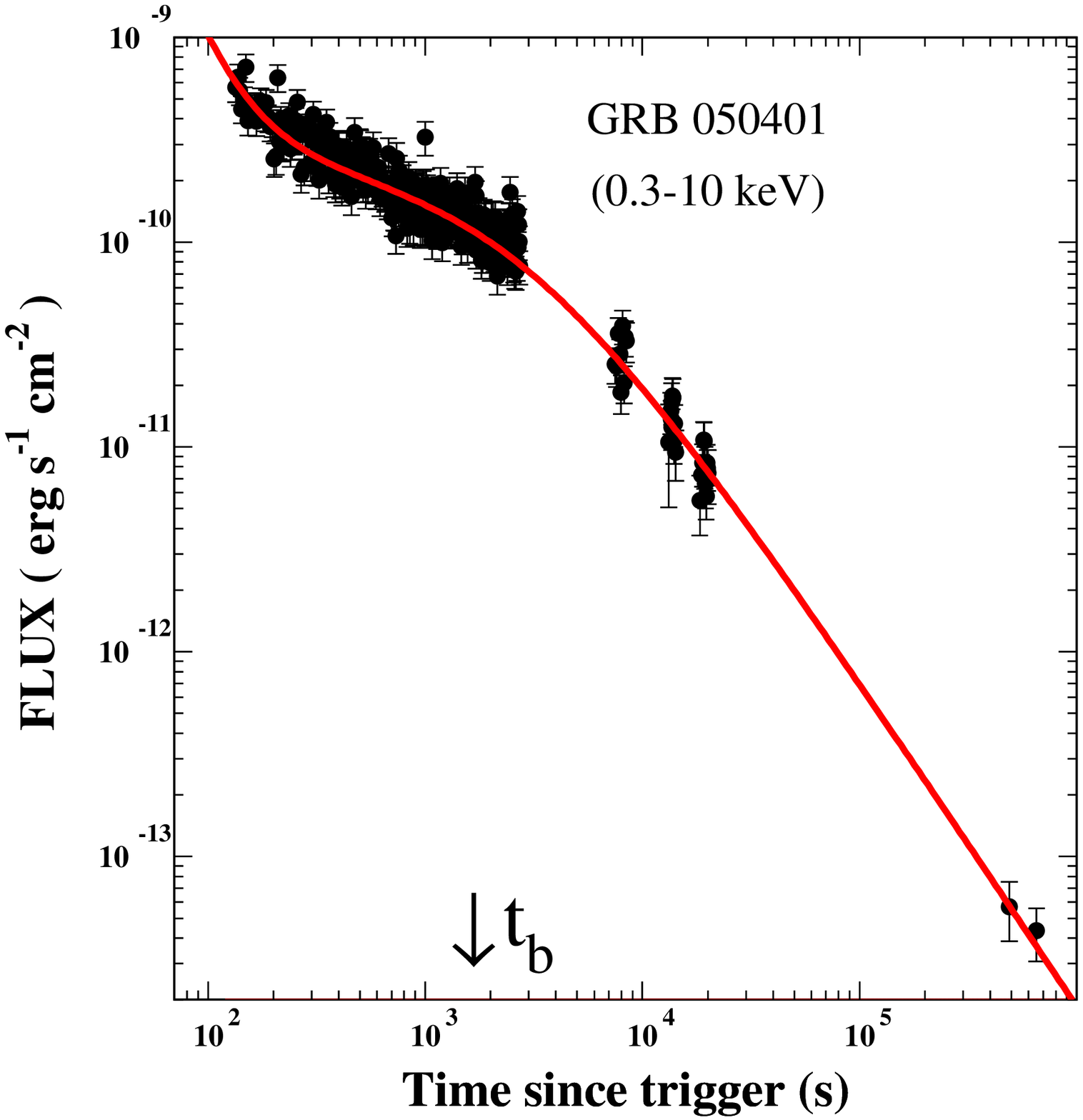,width=8cm} 
}}
\caption{
Comparison between the observed X-ray light curves 
of selected GRBs and their CB model fit: 
{\bf Top left (a):} GRB 980425. The  
last point was measured  1285 days after burst 
(Kouveliotou, et al.~2004). Dotted line: DDD2002.
Continuous line: fit here as all other light curves.
{\bf Top right (b):} GRB 060729.
{\bf Bottom left (c):}  GRB 050505.
{\bf Bottom right (d):} GRB 050401.
All light-curve data, but for GRB 980425,
are from the Swift/XRT light curve repository
(Evans et al.~2007). 
}
\label{f2}
\end{figure}

The CB-model's expectations for the interplay between ICS and SR were
confirmed by the analysis 
of the `canonical' shape of many Swift X-ray light curves (DDD2007a). The 
extreme canonical case is still GRB 980425, shown in Fig.~(\ref{f2}a).
The trend of the `hardness ratios' reflecting the spectral behaviour, and 
the spectral index itself, also corroborate
the expected transitional behaviour (DDD2007b),
as the dominant mechanism evolves from ICS to SR.

\begin{figure}[]
\centering
\vbox{
\epsfig{file=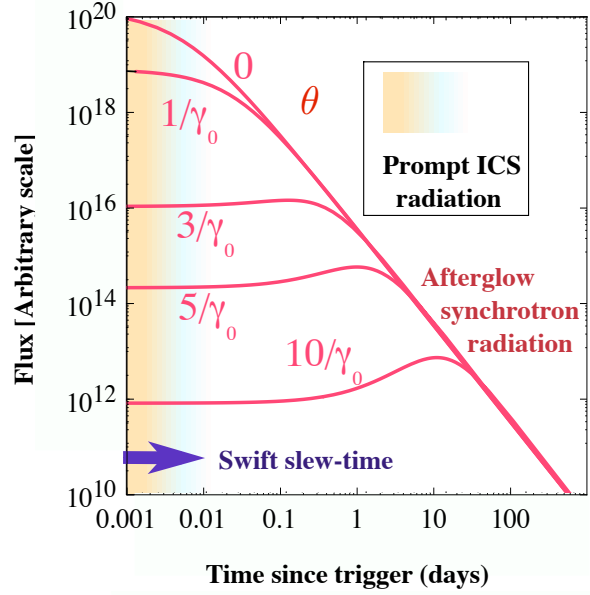,width=8.cm}\\
\epsfig{file=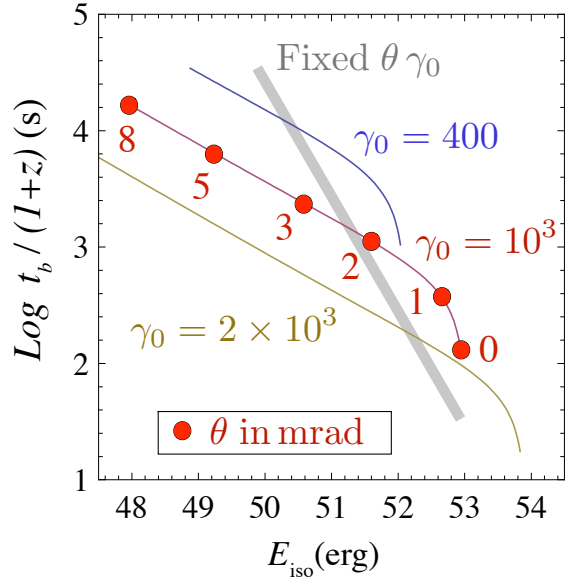,width=7.5cm}
}
\caption{
{\bf Top: (a)} 
Dependence of the synchrotron AG on 
$\theta$, for $\gamma_0\!=\!10^3$ (DDD2002), with $n\,R^2/N_{\rm B}$ as
in Eq.~(\ref{tbreaks}), and a shaded domain for a typical time-zone of
prompt-radiation dominance.
{\bf Bottom: (b)} 
Correlation 
between the  `break-time', $t_b/(1+z)$, of the AG
and the isotropic energy, $E_{\rm iso}$, of the prompt radiation,
for typical parameters and various values of $\gamma_0$.
The dots along the $\gamma_0\!=\! 10^3$ line are labeled with
values of $\theta$ in mrad.
The predicted fast drop of the curves at $\theta\!<\! 1$ mrad is due to the
 CBs not being precisely point-like
(DDD2007d). The thick line is the correlation at $\theta\,\sim\!1/\gamma_0$, the most
probable observer's angle.
}
\label{f1}
\end{figure}

\begin{figure}[]
\centering
\vspace{-1cm}
 \epsfig{file=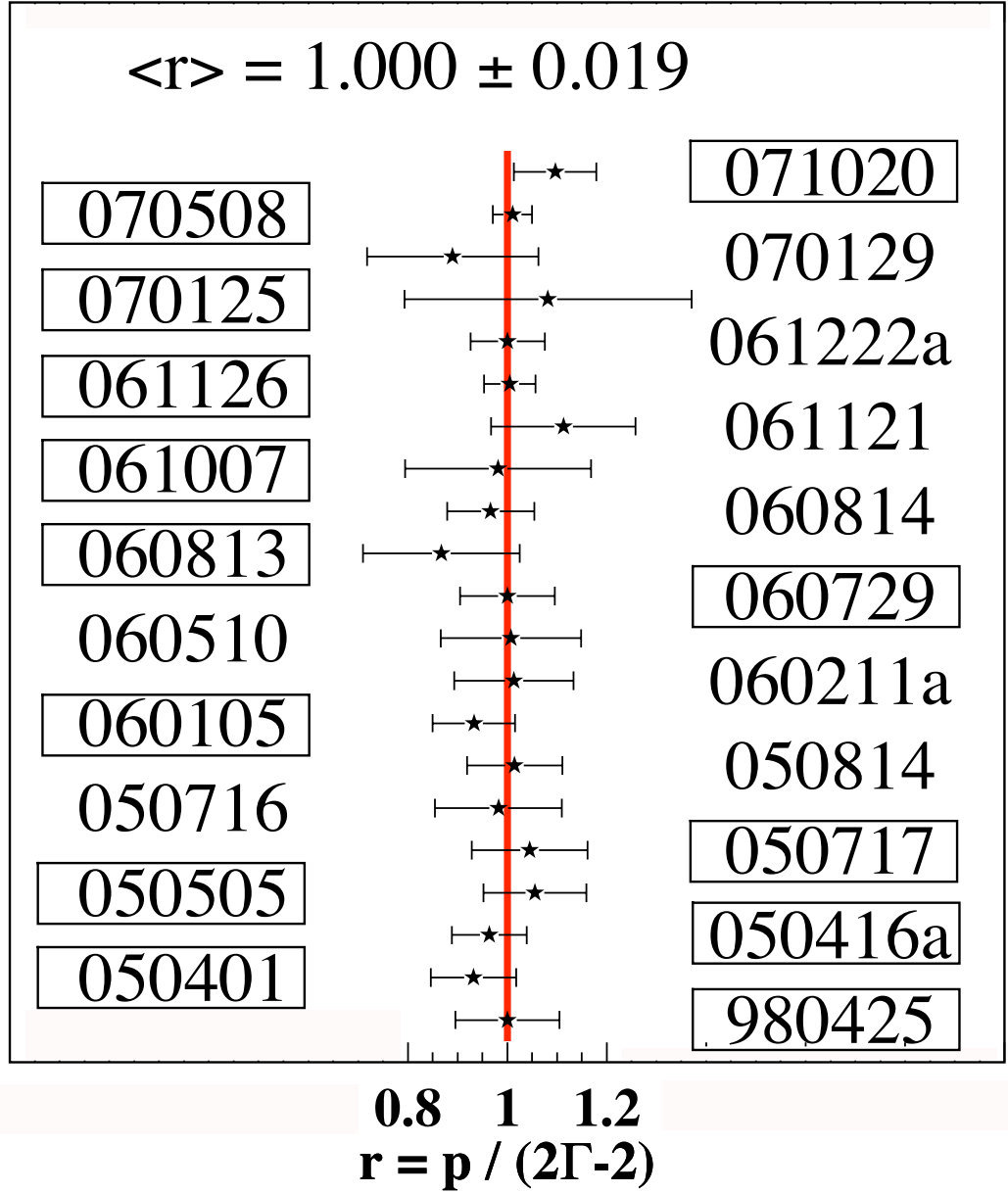,width=18cm}
\caption{
Test of the prediction, $r\!\equiv\! p/(2\Gamma -2)\!=\!1$, of Eq.~(\ref{Asymptotic}),
relating the temporal index, $p$, to the spectral one, $\Gamma$, 
of the afterglows of GRBs. The GRBs discussed in this paper are the outlined ones.
We have extended this test to other GRBs analized in the same fashion.
}
\label{f6}
\end{figure}

In this paper we have shown in detail how the variety of X-ray AG shapes, 
with and without `breaks', is also to be expected from a decelerating jet 
of effectively pointlike cannonballs, as in Fig.~(\ref{f1}a). That the AG 
emission mechanism is SR from CBs slowing down in the way 
approximated by Eq.~(\ref{deceleration})  is confirmed by 
the detailed frequency and time-dependences of Eq.~(\ref{Fnu}) for the 
spectral energy density. We have presented  a study of the 
correlation between the synchrotron AGs' $t$- and $\nu$- dependences, 
specified in the X-ray domain by Eq.~(\ref{Fnux}). This results in a 
relation between the AGs' spectral index $\Gamma$, and the index $p$ 
appearing in their time dependence, a very simple relation at the late times at which 
the time dependence is also a power-law, see Eq.~(\ref{Asymptotic}). The 
prediction is tested in Fig.~\ref{f6}.

\begin{figure}[]
\centering
\vspace{-1cm}
 \epsfig{file=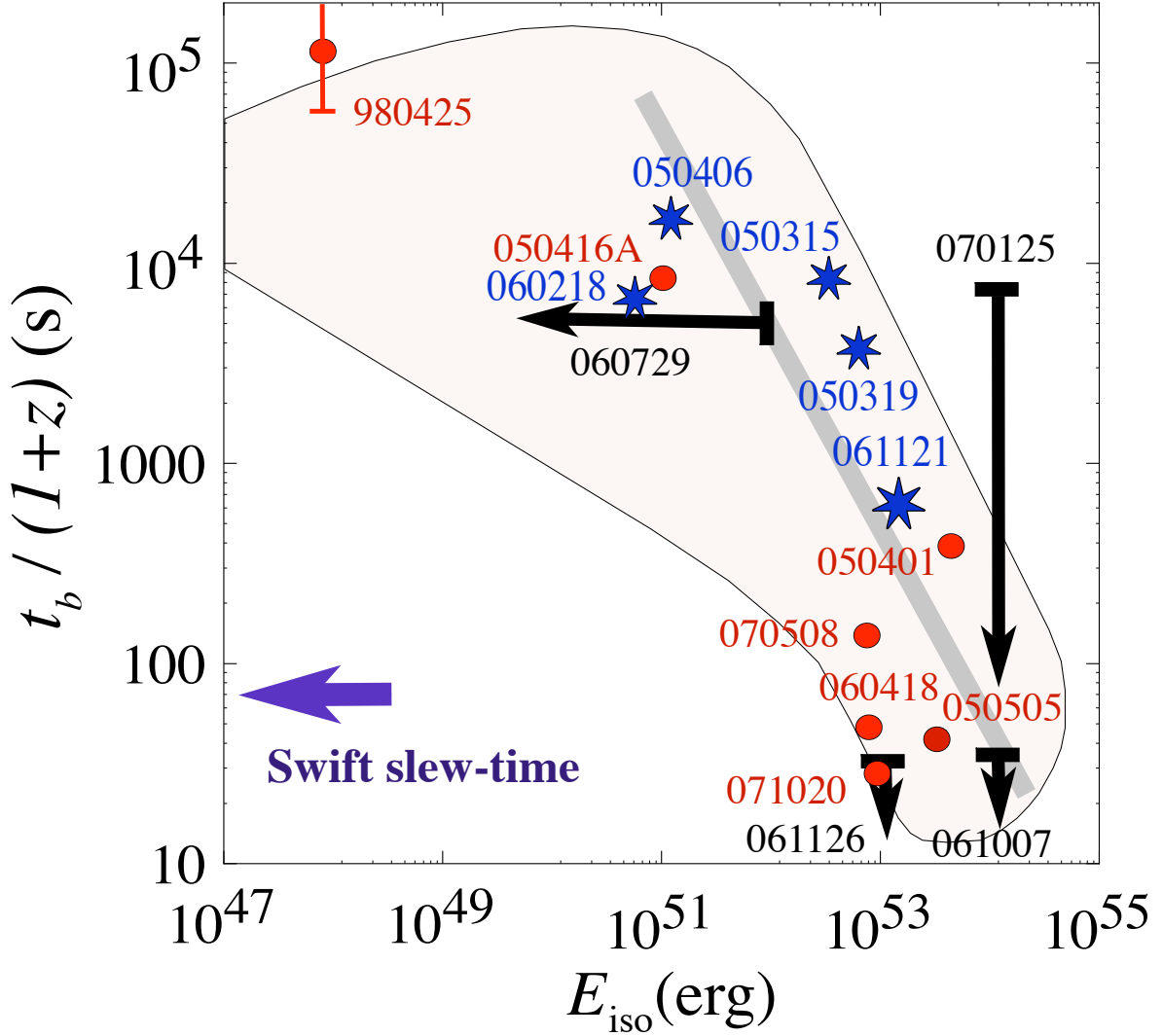,width=16cm}
\caption{
The $[t_b/(1+z),\,E_{\rm iso}]$ correlation. The (red) circles are the GRBs of known 
$E_{\rm iso}$ analized in
this paper, most of which have comparatively small $t_b$. 
The arrows reflect results for which only an upper limit is available.
The (blue) stars are GRBs, mainly with `canonical' X-ray light curves, analized in 
DDD2007a. The large shaded domain is the contour of a region obtained
by letting the parameters vary as specified in the text. The shaded straight line
is the expectation for GRBs viewed close to the most probable angle of observation,
$\theta\,\gamma_0=1$. }
\label{f7}
\end{figure}

\begin{figure}[]
\centering
\vspace{-1cm}
 \epsfig{file=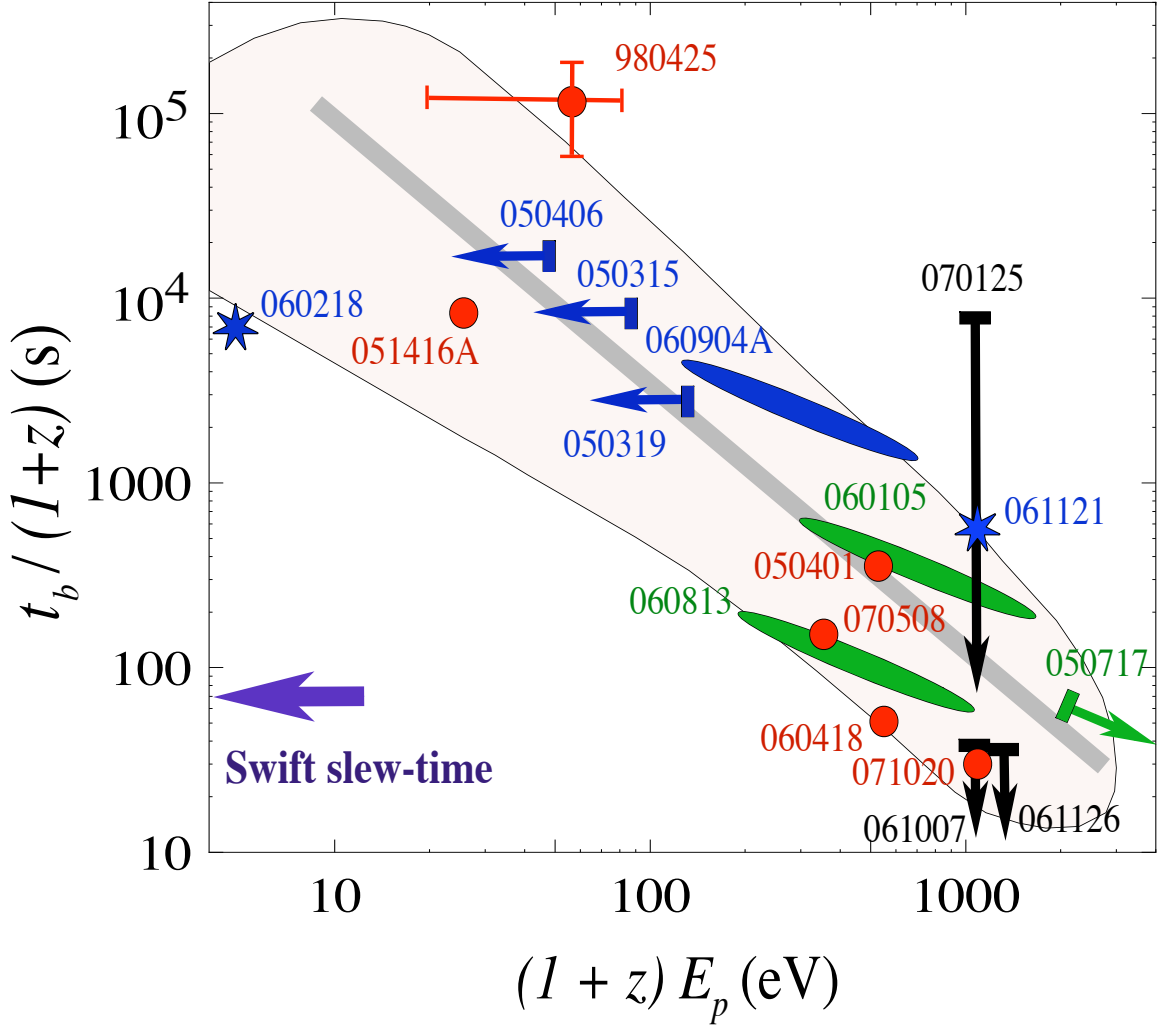,width=16cm}
\caption{
The $[t_b/(1+z),\,E_p]$ correlation. The (red) circles are the GRBs of known $E_p$
analized in this paper, most of which have comparatively small $t_b$. 
The arrows reflect results for which only an upper limit is available.
The ellipses are for GRBs of unknown $z$. The results that are not (red)
circles or vertical arrows
are GRBs, mainly with `canonical' X-ray light curves, analized in 
DDD2007a. The large shaded domain is the contour of a region obtained
by letting the parameters vary as specified in the text. The shaded straight line
is the expectation for GRBs viewed close to the most probable angle of observation,
$\theta\,\gamma_0=1$. The `true' $E_p$ of GRB 980425 could be smaller
than reflected in this plot (Dado \& Dar 2005).
}
\label{f8}
\end{figure}

In the CB model, the understanding of AGs with breaks or no breaks turns
out to be clear: the `missing' SR breaks are hiding under the prompt ICS
radiation, or occur too early to be seen.  This sounds like a trivial and
model-independent excuse. It is not. It is supported by 
our case-by-case analysis of AG shapes. Moreover, a
crucial ingredient ---the angle of observation of the jet,
compared to the beaming angle of its Doppler-boosted radiation--- is
validated by the correlations, e.g.~the luminous AGs are the ones with
early or even undetectable breaks, as in Fig.~\ref{f1}a, and as in many of the
examples we discussed here (the correlation between $t_b$ and the energy
in the X-ray AG was studied in DDD2007d).
Our conclusions are also supported by
the correlations between  the CBs' deceleration-bend  `break-times', $t_b$ 
(in the synchrotron AGs), and the values of $E_{\rm iso}$ 
and $E_p$ (in the prompt Compton signal). 
These correlations, shown in Figs.~\ref{f7} and \ref{f8}, 
reconfirm the consistency of the overall picture.

We have given no comment in the conclusions to our fits to the GRBs and 
XRFs that we have studied. This is because the point we would like 
to make is {\it not} that the CB model can be used to fit the data very 
well. The main issue, in our view, is {\it how} a model, preferably in a 
predictive manner and in terms of very few concrete concepts 
--like its radiation mechanisms, the aperture of its jets and the angle 
from which they are viewed-- can be used to understand the ensemble of long-duration GRBs, 
and XRFs\footnote{XRFs, we allege,  are long-duration GRBs viewed at large angles, 
 but what is a short-duration GRB? We are currently writing a 
paper on how the CB model may shed light on this question.}. After all, 
phenomena that require ever-increasingly complex explanations are of 
limited scientific interest.



\newpage
\begin{deluxetable}{lllllllllc}
\tablewidth{0pt}
\tablecaption{GRB observables and CB-model best-fit afterglow 
parameters.
}
\tablehead{
\colhead{GRB} & \colhead{z}&\colhead{$E_p$}&
\colhead{$E_{\rm iso}$}&
\colhead{$\Gamma$}&
\colhead{p}&  \colhead{$\!\!\!\!\!\!\gamma_0\,\theta$}& \colhead{$t_b[{\rm s}]$}
& \colhead{$t_s[{\rm s}]$} & \colhead{${\chi^2/dof}$} 
}
\startdata

980425 & 0.0085 &  56   & 6.9 E47&
2.1 $\pm$ ?   &  2.20  & 9.17 & 145000 & 36000 & 31/0    \\

060729 & 0.54  &  ---  & $<\!$ 7 E51&
$2.10 \pm 0.15$& 2.20 & 2.51 &    8300 & 130  &  1966/207  \\

050505 & 4.27 &  214 &--- & $1.90\pm 0.20$  &  2.22 & 1.57 &1980 & 
2833 & 114/95   \\

050401 & 2.90  &  132  &3.5 E53 &
$2.18 \pm 0.10$ &2.20   & 0.80   & 1660 & 133  & 353/299  \\

070508 & 0.82 ?   &  188 &7.0 E52 & $2.05\pm 0.04$  &  2.12& 1.22 & 
260 & 82 & 610/469   \\

060813 & ---   &  214 &--- & $1.98\pm 0.18$  &  1.70 & 1.13 &   190 & 
85 & 256/239 \\

060418 & 1.49   & 230   & 9 E52 & $2.03\pm 0.04$    & 2.20 & 1.73  
&  123 & 84  & 339/280  \\

050717 & $>\!$ 2.7 ?   & 2401 & $>\!$ 1 E54 & $1.61\pm 0.10$  & 1.67 & 
(0.08)&  $<\!$ 55 & 91 & 114/78 \\

061126 & 1.159 & 620   & 1.1 E53 &
$1.93 \pm 0.12$ & 1.89  & (1.87)& $<\!$ 104 &1604  & 506/261  \\

071025 & ---   & --- &--- & ---       & 2.20 & (0.90)&  $<\!$ 68& 150& 
330/243 \\

070125 & 1.547 & 440   &9.4 E53& 
$2.10 \pm 0.28$& 2.38& (1.19) & $<\!$ 8060 & 47000 & 28/28 \\

061007 & 1.261 & 498   &1.0 E54&
$2.10 \pm 0.20$ &2.26 & (0.05) & $<\!$ 89 & 86 &  1147/1015  \\

071020 & 2.145 &  322  & 8.0 E52& 
         $1.86\pm 0.06$  & 1.86&    0.67  & 90 & 68 & 234/154  \\
$\;\;\;$ "      &  &      &          &
           & 1.86&  1.43  & 15100 & 68 &  \\

050416A & 0.6535  & 15 &1.2 E51 & $2.04\pm 0.11$  &2.00  & 1.05 &
944 & 85 & 101/92 \\
$\;\;\;$  "    &  &    & &   &  & 2.00  & 
14800 & 85 &    \\

060105 & ---   &  424 &--- & $2.25\pm 0.10$  & 2.33    &0.53   &510& 
96  & 1879/839   \\

060607A & 3.082&---&---&---& 2.20 & 0.97& 164 &   73   & 1119/404  \\

\enddata\\
\vspace{.4cm}
\noindent
{The values of the peak energy, $E_p$ (in keV) 
and $E_{\rm iso}$ (in erg) of the GRBs are from GCN 
reports of
data of Konus-Wind, RHESSI and Suzaku. The GRB redshifts 
are from GCN reports from ground-based optical 
telescopes. The start times $t_s$, of the XRT data after the BAT trigger,
are from the  Swift repository  (Evans et al.~2007).
The unabsorbed spectral indices $\Gamma$ 
 are from Swift GCN reports and Zhang,  Liang \& Zhang (2007). 
 The CB-model fits return $p$, $\gamma_0\,\theta$
 and $t_b$. The parenthesized 
$\gamma_0\,\theta$ 
 are for $t_b$ at its upper limit.}

 \label{t1} 
\end{deluxetable}

\newpage

\newpage

\newpage

\newpage

\newpage

\end{document}